\documentclass[aps,twocolumn,nofootinbib,floatfix,superscriptaddress]{revtex4}
\usepackage{amsfonts}
\usepackage{amssymb}
\usepackage{amsmath}
\usepackage{graphics}
\usepackage{graphicx}
\usepackage{soul}
\usepackage{natbib}
\usepackage{bm}
\usepackage{mathtools}
\usepackage{stmaryrd}
\usepackage{epstopdf}
\usepackage{color}
\usepackage{lipsum}
\usepackage{balance}
\usepackage{enumitem}
\usepackage[colorlinks=true,citecolor=black,linkcolor=black,urlcolor=black,filecolor=black]{hyperref}

\usepackage{acronym}
\newacro{BL}{Balescu--Lenard}
\newcommand{\BL}{\ac{BL}}
\newacro{DF}{distribution function}
\newacroplural{DF}[DFs]{distribution functions}
\newcommand{\DF}{\ac{DF}}

\newacro{HMF}{Hamiltonian Mean Field model}
\newcommand{\HMF}{\ac{HMF}}

\newcommand{\re}{\mathrm{e}}
\newcommand{\ri}{\mathrm{i}}
\newcommand{\rh}{\mathrm{h}}
\newcommand{\rr}{\mathrm{r}}
\newcommand{\p}{\partial}
\newcommand{\rd}{\mathrm{d}}
\newcommand{\rs}{\mathrm{s}}
\newcommand{\mP}{\mathcal{P}}
\newcommand{\Jplus}{J_{\scalebox{0.5}{$+$}}}
\newcommand{\Jminus}{J_{\scalebox{0.5}{$-$}}}
\newcommand{\Dp}{\Delta_{\scalebox{0.5}{$+$}}}
\newcommand{\Dm}{\Delta_{\scalebox{0.5}{$-$}}}
\newcommand{\Gtot}{\Gamma_{\mathrm{tot}}}
\newcommand{\Max}{\mathrm{Max}}
\newcommand{\Min}{\mathrm{Min}}
\newcommand{\eps}{\epsilon}
\newcommand{\veps}{\varepsilon}
\newcommand{\Tdyn}{T_{\mathrm{dyn}}}
\newcommand{\Trelax}{T_{\mathrm{relax}}}
\newcommand{\Ud}{U^{\mathrm{d}}}
\newcommand{\Jp}{J^{\prime}}
\newcommand{\Jpp}{J^{\prime\prime}}

\newcommand{\Omegapp}{\Omega^{\prime\prime}}
\newcommand{\bU}{\mathbf{U}}
\newcommand{\bUd}{\mathbf{U}^{\mathrm{d}}}
\newcommand{\bM}{\mathbf{M}}
\newcommand{\bE}{\mathbf{E}}
\newcommand{\bI}{\mathbf{I}}
\newcommand{\omegaM}{\omega_{\mathrm{M}}}
\newcommand{\omegaR}{\omega_{\mathrm{R}}}
\newcommand{\br}{\mathbf{r}}
\newcommand{\brp}{\mathbf{r}^{\prime}}
\newcommand{\thetap}{\theta^{\prime}}
\newcommand{\rp}{r^{\prime}}
\newcommand{\rmin}{r_{\mathrm{min}}}
\newcommand{\rmax}{r_{\mathrm{max}}}
\newcommand{\half}{\tfrac{1}{2}}
\newcommand{\deltaD}{\delta_{\mathrm{D}}}

\newcommand{\Htot}{H_{\mathrm{tot}}}
\newcommand{\Uext}{U_{\mathrm{ext}}}
\newcommand{\Fb}{F_{\mathrm{b}}}
\newcommand{\Jb}{J_{\mathrm{b}}}
\newcommand{\Gtotb}{\Gamma_{\mathrm{tot}}^{\mathrm{b}}}
\newcommand{\sigmab}{\sigma_{\mathrm{b}}}
\newcommand{\Omegab}{\Omega_{\mathrm{b}}}
\newcommand{\Etot}{E_{\mathrm{tot}}}
\newcommand{\Ltot}{L_{\mathrm{tot}}}
\newcommand{\ext}{\mathrm{ext}}
\newcommand{\self}{\mathrm{self}}
\newcommand{\bJ}{\mathbf{J}}
\newcommand{\bO}{\mathbf{\Omega}}
\newcommand{\bk}{\mathbf{k}}
\newcommand{\Lambdatot}{\Lambda_{\mathrm{tot}}}
\newcommand{\Lambdad}{\Lambda^{\mathrm{d}}}
\newcommand{\mU}{\mathcal{U}}
\newcommand{\mR}{\mathcal{R}}
\newcommand{\Treg}{T_{\mathrm{reg}}}
\newcommand{\Tnl}{T_{\mathrm{nl}}}
\newcommand{\kmax}{k_{\mathrm{max}}}
\newcommand{\Fd}{F_{\mathrm{d}}}
\newcommand{\Hd}{H_{\mathrm{d}}}
\newcommand{\mO}{\mathcal{O}}
\newcommand{\Tmax}{T_{\mathrm{max}}}
\newcommand{\Nreal}{N_{\mathrm{real}}}
\newcommand{\mF}{\mathcal{F}}
\newcommand{\FB}{F_{\mathrm{B}}}
\newcommand{\bX}{\mathbf{X}}
\newcommand{\bY}{\mathbf{Y}}
\newcommand{\msJ}{\mathsf{J}}
\newcommand{\mD}{\mathcal{D}}
\newcommand{\Jstar}{J_{\star}}
\newcommand{\Dstar}{D_{\star}}
\newcommand{\Fstar}{F_{\star}}
\newcommand{\rTN}{\mathrm{TN}}
\newcommand{\Jr}{J_{\mathrm{r}}}
\newcommand{\Sign}{\mathrm{Sign}}
\newcommand{\iso}{\mathrm{iso}}
\newcommand{\mDsol}{\mathcal{D}_{\mathrm{sol}}}

\newcommand{\Jtest}{J_{\mathrm{test}}}
\newcommand{\tmin}{t_{\mathrm{min}}}
\newcommand{\tmax}{t_{\mathrm{max}}}
\newcommand{\Dnum}{D_{\mathrm{num}}}
\newcommand{\Jsep}{J_{\mathrm{sep}}}
\newcommand{\Tregc}{T_{\mathrm{reg}}^{\mathrm{c}}}
\newcommand{\Dstarc}{D_{\star}^{\mathrm{c}}}
\newcommand{\Dstart}{D_{\star}^{\mathrm{t}}}

\usepackage{soul} \usepackage{xcolor}

\begin{document}

\title{Kinetic theory of two-dimensional point vortices at order ${1/N}$ and ${1/N^{2}}$}

\author{Jean-Baptiste Fouvry}
\affiliation{Institut d'Astrophysique de Paris, UMR 7095, 98 bis Boulevard Arago, F-75014 Paris, France}
\author{Pierre-Henri Chavanis}
\affiliation{Laboratoire de Physique Th\'eorique, Universit\'e de Toulouse, CNRS, UPS, France}

\begin{abstract}
We investigate the long-term relaxation of a distribution of $N$ point vortices
in two-dimensional hydrodynamics.
To focus on the regime of weak collective amplification,
we embed these point vortices within a static background potential
and soften their pairwise interaction on small scales.
Placing ourselves within the limit of an average axisymmetric distribution,
we stress the connections with generic long-range interacting systems,
whose relaxation is described within angle-action coordinates.
In particular, we emphasise the existence of two regimes of relaxation,
depending on whether the system's profile of mean angular velocity (frequency)
is a non-monotonic [resp.\ monotonic] function of radius,
which we refer to as profile (1) [resp.\ profile (2)].
For profile (1), relaxation occurs 
through two-body non-local resonant couplings, i.e.\ $1/N$ effects,
 as described by the inhomogeneous Landau equation.
For profile (2), the impossibility of such two-body resonances
submits the system to a ``kinetic blocking''. Relaxation is then driven by three-body couplings,
i.e.\ ${1/N^{2}}$ effects, whose associated kinetic equation has only recently been derived.
For both regimes, we compare extensively the kinetic predictions
with large ensemble of direct $N$-body simulations.
In particular, for profile (1),
we explore numerically an effect akin to ``resonance broadening''
close to the extremum of the angular velocity profile.
Quantitative description of such subtle nonlinear effects
will be the topic of future investigations.
\end{abstract}
\maketitle

\section{Introduction}
\label{sec:Introduction}

Two-dimensional (${2D}$) geophysical and astrophysical flows have the striking
property of organising spontaneously into coherent structures (large-scale
vortices)~\cite{Chavanis2002,Bouchet+2012}. A famous example of this
self-organization is Jupiter's great red spot. Since these coherent structures
result from a turbulent mixing, it has sometimes been said that ``order emerges
from chaos''. To account for this self-organization, Onsager~\cite{Onsager1949}
considered the statistical mechanics of a system of point vortices of
circulations $\gamma_{i}$ (${ i \!=\! 1,...,N }$) on the plane. He showed the existence of
negative temperature states\footnote{This was a few years before the discovery
of negative temperature states in nuclear spin systems~\cite{Purcell+1951}.
In his visionary paper, Onsager also predicted, in a
famous footnote, the quantization of the circulation of point vortices in units
of ${ h / m }$ in superfluid helium.} at which point vortices of the same sign tend to
group themselves into ``supervortices''. His qualitative arguments were later
developed by Joyce and Montgomery~\cite{Joyce+1973,Montgomery+1974}, Kida~\cite{Kida1975}
and Pointin and Lundgren~\cite{Pointin+1976,Lundgren+1977} in a mean field approximation valid for ${ N \!\to\! + \infty }$,
with ${ \gamma \!\sim\! 1/N }$.
Overall, this leads to the Boltzmann distribution coupled
to the Poisson equation.\footnote{In unpublished notes~\cite{Eyink+2006}, Onsager had
previously developed this mean field theory by analogy with his research on
electrolytes, which are also described by the Boltzmann--Poisson
equations.} This
equilibrium state (most probable state) can be obtained by maximising the
Boltzmann entropy at fixed circulation, energy, and angular momentum. Later,
Miller~\cite{Miller1990}, as well as Robert and Sommeria~\cite{Robert+1991},
considered the statistical mechanics of more realistic continuous vorticity fields (instead of
singular point vortices) described by the ${2D}$ Euler--Poisson equations.
In particular, they derived
a generalization of the Boltzmann distribution involving a continuum of
vorticity levels. For a long time, the distinction between these two statistical
theories was not fully understood and appreciated. It was clarified only later by
developing the analogy between ${2D}$ vortices and stellar systems~\cite{Chavanis+1996,Chavanis2002}.

Indeed, self-gravitating systems also self-organize into coherent 
structures like star clusters, galaxies, or even clusters of galaxies~\cite{Binney+2008}.
 Stellar systems, which are fundamentally discrete (i.e.\ made of a collection of
$N$ stars in gravitational interaction), undergo two successive types of
evolution. In a first regime, one can ignore gravitational encounters between
stars and make a mean field approximation. In this collisionless regime, the
locally averaged distribution function of stellar systems is governed by the
Vlasov--Poisson equations. Collisionless stellar systems generically experience a
process of violent relaxation towards a quasistationary (virialised) state
which is a stable stationary solution of the Vlasov--Poisson equations.
Lynden-Bell~\cite{LyndenBell1967} tried to predict this quasistationary state in terms of
statistical mechanics but his prediction is spoiled by the problem of incomplete
relaxation~\cite{LyndenBell1967,Chavanis2006}. Then, on longer timescales,
stellar systems evolve under the effect of gravitational encounters (finite-$N$ effects,
granularities, correlations...). During this collisional regime, stellar
systems have the 
tendency to become isothermal but their relaxation towards a true statistical
equilibrium state described by the Boltzmann--Poisson equation\footnote{Actually,
there is no statistical equilibrium state for self-gravitating systems in ${3D}$
because the Boltzmann distribution coupled to the Poisson equation has an
infinite mass. Said differently, there is no maximum entropy state at fixed mass
and energy.} is hampered by physical effects such as the evaporation of high
energy stars and the gravothermal catastrophe (core collapse)~\cite{LyndenBell+1968}. 

Similar results apply to ${2D}$ vortices. A gas of point vortices, like a gas 
of stars, undergoes two successive types of evolution. In a first regime, one
can ignore finite-$N$ effects and make a mean field approximation. In this
collisionless limit, the locally averaged vorticity of ${2D}$ point vortices is
governed by the ${2D}$ Euler--Poisson equations. Collisionless point vortices
generically experience a process of violent relaxation towards a quasistationary
state which is a stable stationary solution of the ${2D}$ Euler--Poisson equations.
The Miller--Robert--Sommeria~\cite{Miller1990,Robert+1991} statistical theory in fluid
mechanics, which applies to this regime, is the counterpart of the Lynden-Bell theory~\cite{LyndenBell1967}
in stellar dynamics. Its predictions are also affected by the
problem of incomplete relaxation. Then, on longer timescales (collisional
regime), ${2D}$ point vortices, just like stellar systems,
evolve under the effect of ``collisions''
towards the statistical equilibrium state described
by the Boltzmann--Poisson equation~\cite{Joyce+1973,Montgomery+1974,Kida1975,Pointin+1976,Lundgren+1977}.

One can then develop the kinetic theory of ${2D}$ point vortices by analogy 
with the kinetic theory of stellar systems pioneered
by Chandrasekhar~\cite{Chandrasekhar1942,Chandrasekhar1943a,Chandrasekhar1943b,Chandrasekhar1949}. By making a thermal bath approximation
and using the methods of Brownian theory~\cite{Chandrasekhar1943Brownian}, Chandrasekhar
showed that a test star experiences a stochastic process governed by a
Fokker--Planck equation involving a diffusion term and a friction term.
Similarly, under the same assumptions, a test vortex experiences a stochastic
process governed by a Fokker--Planck equation involving a diffusion term and a
drift term~\cite{Chavanis1998,Chavanis2001}. As such, the drift of a point vortex is the counterpart
of the dynamical friction undergone by a star. These Fokker--Planck equations
relax both towards the Boltzmann distribution. In addition, the diffusion and friction (or drift)
coefficients are related to each other and to the temperature through
some Einstein relation. In particular, in the case of point vortices,
the temperature may be negative.

However, this thermal bath approach is oversimplified and does not correctly 
describe the collisional evolution of the system as a whole. In particular, the
Fokker--Planck equation does not conserve the total energy.
A better kinetic equation for systems with
long-range interactions is the Landau equation~\cite{Landau1936} or the
Balescu--Lenard equation~\cite{Balescu1960,Lenard1960} which were originally introduced for homogeneous
plasmas and applied in a slightly different form to stellar
systems by making a local approximation~\cite{Rosenbluth+1957}.
These kinetic equations take into
account two-body correlations (collisions), so they are valid at order ${ 1/N }$.
In the homogeneous case, they relax towards the Maxwell-Boltzmann distribution on
a timescale ${ (N/\ln N)\, \Tdyn }$, due to logarithmic corrections. These kinetic
equations were later extended to describe the axisymmetric evolution of a
point vortex gas as a whole~\cite{Dubin+1988,Dubin2003,Chavanis2001,Chavanis+2007,Chavanis2008,Chavanis2010,Chavanis2012Vortex,Chavanis2012Onsager,Fouvry+2016,Chavanis2023}. It turns out that their structure
is seemingly different from the original (homogeneous) Landau and Balescu--Lenard
equations. Indeed, they involve a resonance condition on the
profile of angular velocity ${ \Omega(r,t) }$. As long as the profile of angular
velocity is non-monotonic, the resonance condition can be satisfied and the
system relaxes through two-body effects.
However, the relaxation stops when the profile of
angular velocity becomes monotonic, even if the system has not reached the
statistical equilibrium state.
This is an example of ``kinetic blocking''~\cite{Chavanis2001,Chavanis+2007}.
A similar situation of kinetic blocking had been previously
observed for ${1D}$ homogeneous plasmas~\cite{Eldridge+1962}.
In that case, the Balescu--Lenard
collision term vanishes identically and the evolution of the system is governed
by a collision term of order ${ 1/N^{2} }$ that accounts for three-body correlations.
An explicit expression for this collision term has been obtained recently in the case of
homogeneous ${1D}$ systems with long-range interactions,
when collective effects are neglected~\cite{Fouvry+2019,Fouvry+2020}.
It always drives the relaxation towards the
Boltzmann distribution, showing that the relaxation time scales as ${ N^2 \Tdyn }$
in that case.

In reality, stellar systems are spatially inhomogeneous.
Their kinetic theory is
described by the inhomogeneous Balescu--Lenard equation written with angle-action
variables~\cite{Heyvaerts2010,Chavanis2012AA}.
When collective effects are neglected
(i.e.\ for sufficiently hot systems),
it reduces to the inhomogeneous
Landau equation~\cite{Chavanis2013}. These equations involve a resonance condition
that depends on the orbital frequencies, ${ \bO (\bJ , t) }$,
which themselves are functions of the actions, $\bJ$.
This is similar to the kinetic equation for ${2D}$ point vortices with more general
variables.
For ${1D}$ inhomogeneous long-range interacting systems with a monotonic
frequency profile in which only ${ 1\!:\!1 }$ resonances are permitted,
there can also be a kinetic blocking, just like for axisymmetric point vortices.
In that case, one has to develop the kinetic theory at the order ${ 1/N^2 }$. The
expression of the collision term in the absence of collective effects has been
obtained in~\cite{Fouvry2022}. Except for very special interaction potentials~\cite{Fouvry+2023}, this equation drives the system towards the Boltzmann distribution on a timescale ${ N^{2} \, \Tdyn }$.

Building upon this rich history of parallel theoretical developments,
in the present work, we leverage the generality
of the kinetic theory of (inhomogeneous) long-range interacting systems
to apply it to describe the long-term relaxation of axisymmetric
distribution of point vortices.

First, we focus on a case with a non-monotonic distribution
of orbital frequency (i.e.\ the vortices' angular velocity) [profile (1)].
To place ourselves in the limit where collective effects can be neglected,
we embed the point vortices within a static background potential
and soften the pairwise interaction on small scales.
In that case, relaxation can be described with the inhomogeneous Landau equation.
We compare quantitatively its prediction with measurements
in direct $N$-body simulations, focusing in particular
on the dependence of the relaxation rate with the total number of particles
and their active fraction (i.e.\ the fraction of self-consistency in the mean potential).
Importantly, we unveil how, close to the extremum of the angular velocity profile,
the effective relaxation in the $N$-body simulations is reduced compared to the kinetic prediction.
We attribute this effect to ``resonance broadening'' through nonlinear effects
and present a heuristic correction to the Landau equation
that closely matches the numerical measurements.
Surprisingly, we show that as one delays the impact of nonlinearities,
e.g.\@, by increasing the number of particles and/or lowering the active fraction,
the resonance broadening seems to saturate
and the convergence toward the Landau prediction halts.

Second, we focus on the case of an axisymmetric distribution of point vortices
with a monotonic distribution of orbital frequency.
In that case, both the ${ 1/N }$ Landau and ${ 1/N }$ Balescu--Lenard equations
predict a vanishing flux: a ``kinetic blocking''~\citep[see, e.g.\@,][]{Chavanis2001,Chavanis+2007}.
Once again, we place ourselves in the limit of weak collective amplification
by embedding the point vortices within a static background potential
and by softening their pairwise interaction on small scales.
In that dynamically hot regime,
we compare quantitatively measurements of the relaxation rate in numerical simulations
with the prediction from the recently derived ${1/N^2}$ Landau equation~\cite{Fouvry2022}.
We obtain a satisfying agreement between both,
recovering in particular the appropriate dependence
with respect to the total number of particles and their active fraction.

The paper is organised as follows.
In Section~\ref{sec:System}, we detail our system.
In Section~\ref{sec:1_N_dynamics}, we investigate
long-term dynamics as driven by ${1/N}$ effects [profile (1)],
while in Section~\ref{sec:1_N2_dynamics},
focusing on systems undergoing a kinetic blocking [profile (2)],
we consider ${1/N^{2}}$ effects.
We conclude in Section~\ref{sec:Conclusion}.
Throughout the main text, technical details are kept to a minimum
and deferred to Appendices or to relevant references.

\section{System}
\label{sec:System}

\subsection{Hamiltonian}
\label{sec:Hamiltonian}

We consider a ${2D}$ system composed of $N$ point vortices
of individual circulation ${ \gamma_{i} \!=\! \Gtot / N }$,
with $\Gtot$ the system's total active circulation.
The dynamics of these vortices is given by the Kirchhoff--Hamilton equations~\citep{Newton2001}
\begin{equation}
\gamma_{i} \, \frac{\rd x_{i}}{\rd t} = \frac{\p \Htot}{\p y_{i}} ,
\qquad
\gamma_{i} \, \frac{\rd y_{i}}{\rd t} = - \frac{\p \Htot}{\p x_{i}} ,
\label{eq:EOM_generic}
\end{equation}
with ${ (x_{i} , y_{i}) \!=\! \br_{i} }$ the location of the point vortex $i$.
Here, as visible in Eq.~\eqref{eq:EOM_generic},
$x_{i}$ and $y_{i}$ should be interpreted
as canonically conjugate.
The system's total Hamiltonian is
\begin{equation}
\Htot = \sum_{i < j} \gamma_{i} \gamma_{j} \, U (\br_{i} , \br_{j}) ,
\label{eq:def_Htot}
\end{equation}
with the pairwise interaction
\begin{equation}
U (\br , \brp) = - \frac{G}{2 \pi} \, \ln |\br \!-\! \brp| 
\label{eq:def_U_unsoftened}
\end{equation}
of typical amplitude $G$.
We point out that $\Htot$ does not contain any kinetic energy term
in the usual sense.
We also recall that the $2D$ Newtonian potential $U$
from Eq.~\eqref{eq:def_U_unsoftened} satisfies the Poisson equation
${ \Delta_{\br} U(\br,\brp) \!=\! - G \, \deltaD (\br \!-\! \brp) }$.

In practice, to cure possible divergences on small scales in Eq.~\eqref{eq:def_U_unsoftened},
we replace the pairwise interaction with its softened analog
\begin{equation}
U_{\eps} (\br , \brp) = - \frac{G}{2 \pi} \ln \big( \big[ |\br \!-\! \brp|^{2} \!+\! \eps^{2} \big]^{1/2} \big) ,
\label{eq:def_U_softened}
\end{equation}
with $\eps$ some (small) softening length.
This greatly eases the implementation of the $N$-body simulations
(see Appendix~\ref{app:NumericalSimulations}).

\subsection{Axisymmetric limit}
\label{sec:Axisymmetric}

\begin{figure}
\begin{center}
\raisebox{-0.5\height}{\includegraphics[width=0.23\textwidth]{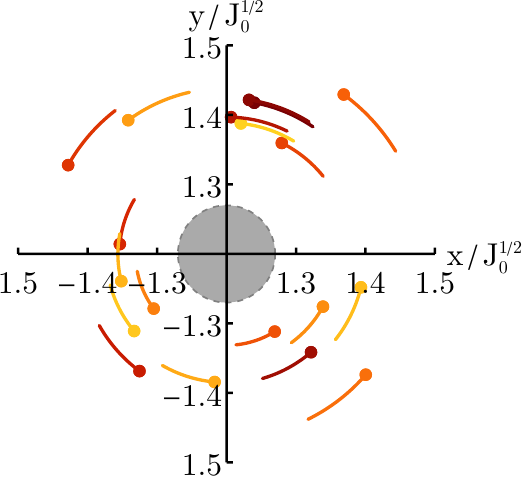}}
\raisebox{-0.5\height}{\includegraphics[width=0.23\textwidth]{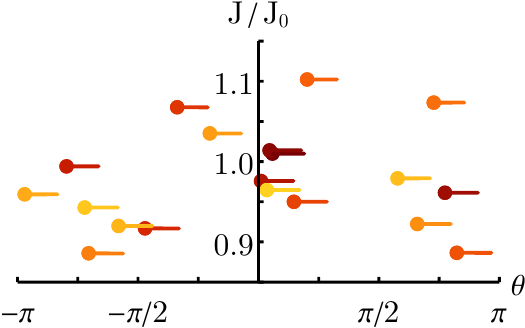}}
\caption{Left: Illustration in configuration space, ${ (x,y) }$,
of the dynamics of a few point vortices following the \DF\ from Eq.~\eqref{eq:DF_LOC}.
Right: Same, but in action space ${ (\theta , J) }$. In that space,
the vortices follow (perturbed) straight-line trajectories.
\label{fig:System}}
\end{center}
\end{figure}

When averaged over independent realisations,
the system's statistics is described via its \DF\@, ${ F \!=\! F (\br , t) }$,
normalised to ${ \!\int\! \rd \br F \!=\! \Gtot }$.
We assume that the mean DF, ${ F({\bf r},t) \!=\! \langle \sum_{i} \gamma_i \deltaD [\br \!-\! \br_{i}(t) ] \rangle }$ is axisymmetric so that one has ${ F \!=\! F (r , t) }$ with ${ r \!=\! |\br| }$. Subsequently, the mean individual Hamiltonian ${ H(\br , t) \!=\! \int \rd \brp U(\br , \brp) F (\brp, t) }$ is ${ H \!=\! H (r , t) }$.
We can then introduce appropriate angle-action coordinates
${ (\theta , J) }$,
with $\theta$ the ${ 2\pi }$-periodic angle
and $J$ the associated action,
via\footnote{These follow from the generating function ${ S(y,\theta) \!=\! \half y^{2} \tan \theta }$, to be used in eq.~{(D.94)} of~\cite{Binney+2008}.}
\begin{equation}
\theta = \tan^{-1} (x / y) ,
\qquad
J = \half r^{2} .
\label{eq:def_AA}
\end{equation}
Importantly, volume conservation ensures that ${ \rd \br \!=\! \rd \theta \rd J }$.
With such a choice, it is now clear how an axisymmetric distribution
of point vortices 
can simply be interpreted as one example of
a long-range interacting inhomogeneous systems~\citep[see, e.g.\@,][]{Chavanis2012AA},
characterised by some mean \DF\@, ${ F \!=\! F (J , t) }$.
Along the same line, from the mean Hamiltonian, ${ H \!=\! H (J,t) }$,
we define the mean orbital frequency, ${ \Omega (J,t) \!=\! \p H / \p J }$.\footnote{In the original point vortex model~\citep[see, e.g.\@,][]{Chavanis2012Vortex}, the DF ${ F(\br,t) }$ corresponds to the vorticity ${ \omega(\br, t) }$,
the individual Hamiltonian ${ H(\br , t ) }$ corresponds to the stream function
${ \psi(\br , t ) }$, and the orbital frequency ${ \Omega(J,t) }$
corresponds to the angular velocity ${ \Omega(r,t) }$ (for axisymmetric flows).
For the $2D$ Newtonian interaction potential,
one has ${ \omega \!=\! - \Delta \psi }$
and ${ \Omega(r,t) \!=\! \frac{1}{r^2} \int_{0}^{r} \rd \rp \, \rp \omega(\rp,t) }$.}
In Fig.~\ref{fig:System},
we illustrate the typical evolution in configuration space
and action space of such axisymmetric distributions.
The goal of kinetic theory is to describe the mean long-term evolution
of the orbital distribution, i.e.\ to predict ${ \p F (J , t) / \p t }$.
Let us now describe the setup considered in this work.

\subsection{Initial conditions}
\label{sec:InitialConditions}

\begin{figure}
\begin{center}
\includegraphics[width=0.45\textwidth]{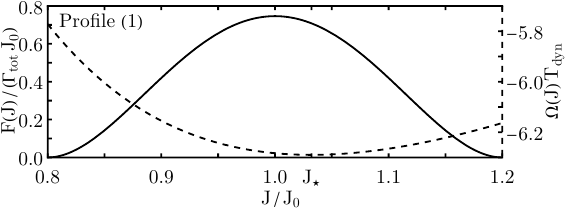}
\includegraphics[width=0.45\textwidth]{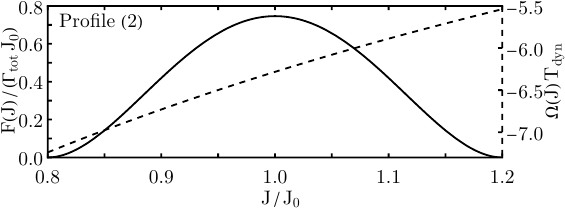}
\caption{Illustration of the \DF\ from Eq.~\eqref{eq:DF_LOC} (full lines)
and the considered frequency profiles (dashed lines).
Profile (1) (top panel) is non-monotonic and has an extremum in ${ \Jstar \!=\! 1.03 \, J_{0} }$,
while profile (2) (bottom panel) is monotonic.
\label{fig:DF}}
\end{center}
\end{figure}

For the \DF\@, we consider an initial condition of the form
\begin{equation}
F (J) \!=\! A \bigg( \frac{J \!-\! J_{0}}{\sigma_{0}} \!+\! 1 \bigg)^{2} \bigg( \frac{J \!-\! J_{0}}{\sigma_{0}} \!-\! 1 \bigg)^{2} \! \Theta \big[ |J \!-\! J_{0}| \!\leq\! \sigma_{0} \big] ,
\label{eq:DF_LOC}
\end{equation}
with ${ \Theta [x \!\geq\! 0] }$ the usual Heaviside function
and ${ A \!=\! 15 \Gtot / (32 \pi \sigma_{0}) }$
to ensure ${ \!\int\! \rd J \rd \theta F \!=\! \Gtot }$.
As illustrated in Fig.~\ref{fig:DF},
this \DF\ is characterised by two parameters,
namely $J_{0}$ its central location
and $\sigma_{0}$ its width.
This \DF\ is non-zero only within the domain ${ |J \!-\! J_{0}| \!\leq\! \sigma_{0} }$
and is continuous, along with its first derivative,
on the entire action domain.
As such, it is a convenient \DF\ with which to explore
long-term relaxation.

\subsection{Active fraction}
\label{sec:ActiveFraction}

To limit ourselves to the dynamically hot limit,
i.e.\ the limit of small amplification of perturbations,
we assume that part of the Hamiltonian
is not self-consistently generated by the vortices.
As such, we introduce an active fraction, ${ 0 \!<\! q \!<\! 1 }$,
and make the replacement
\begin{equation}
\gamma_{i} \to q \, \gamma_{i} ,
\label{eq:intro_q}
\end{equation}
so that the total circulation ${ \Gtot \!=\! \sum_{i=1}^{N} \!\gamma_{i} }$
is also reduced by a factor $q$.
Following this modification,
we add an external potential to our system
to ensure that systems with different $N$ and $q$
share the exact same mean-field orbital frequencies.

More precisely, we replace the total Hamiltonian
from Eq.~\eqref{eq:def_Htot} with
\begin{equation}
\Htot = \sum_{i = 1}^{N} \gamma_{i} \, \Uext (\br_{i}) + \sum_{i < j}^{N} \gamma_{i} \, \gamma_{j} \, U_{\eps} (\br_{i} , \br_{j}) .
\label{eq:def_Htot_tambouille}
\end{equation}
In that expression, $U_{\eps}$ is the softened pairwise interaction
from Eq.~\eqref{eq:def_U_softened}.
We also introduced an external potential, $\Uext$, that reads
\begin{equation}
\Uext = H_{0} - H_{\eps} [F].
\label{eq:def_Uext}
\end{equation}
In that expression, ${ H_{\eps} [F] (\br) \!=\! \! \int \! \rd \brp F(\brp,t\!=\!0) \, U_{\eps} (\br , \brp) }$
is the (softened) potential generated by the initial \DF\
from Eq.~\eqref{eq:DF_LOC}.
We also introduced ${ H_{0} \!=\! H_{0} (J) }$ an axisymmetric external potential.
Importantly, with the present choice,
at ${ t \!=\! 0 }$, we have ${ \langle \Htot \rangle \!=\! \sum_{i} \gamma _{i} \, H_{0} (J) }$, i.e.\ the initial mean-field orbital frequencies
are simply ${ \Omega (J) \!=\! \rd H_{0} / \rd J }$.
From it, we also define the dynamical time as ${ \Tdyn \!=\! 2 \pi / |\Omega (J_{0})| }$.
With this setup, changing the value of $N$ amounts to changing
the amplitude of the Poisson fluctuations,
while changing $q$ amounts to changing the dynamical temperature,
i.e.\ the level of self-consistent amplification in the system.
Indeed, the smaller $q$, the less the point vortices
contribute themselves to the overall total Hamiltonian.
Importantly, changing $N$ or changing $q$ does not lead to any change
in the mean frequency profile. In the following, for simplicity, we assume that the point vortices have the same circulation, ${ \gamma_{i} \!=\! \gamma }$.

\subsection{Frequency profiles}
\label{sec:FrequencyProfiles}

For the frequency profiles,
we consider two different cases,
profiles (1) and (2), as illustrated in Fig.~\ref{fig:DF}.
As detailed in Appendix~\ref{app:BackgroundProfiles},
these correspond to the (unsoftened) potential
generated by a simple axisymmetric background distribution (Eq.~\ref{eq:DF_Profile_1}).\footnote{We emphasise that the orbital frequency (angular velocity) profile ${ \Omega(J) }$ is not self-consistently generated by the point vortices that constitute our system,
i.e.\ by the \DF\ from Eq.~\eqref{eq:DF_LOC}, but is imposed from the outside,
e.g.\@, by being generated by a background distribution of point vortices exterior to our system.
This allows us to prescribe the orbital frequency in a convenient manner.}
The main difference between these two frequency profiles,
${ J \!\mapsto\! \Omega(J) }$,
is that profile (1) is non-monotonic,
exhibiting a minimum in ${ \Jstar \!\simeq\! 1.03 \, J_{0} }$,
while profile (2) is monotonic.
As the upcoming section will detail,
this difference plays a crucial role in defining the system's regime of relaxation.

\section{${ 1/N }$ dynamics}
\label{sec:1_N_dynamics}

\subsection{Kinetic equation}
\label{sec:1N_kinetic_equation}

First, let us consider profile (1).
At order ${1/N}$, relaxation is driven by two-body effects.
In the limit where collective effects can be neglected,
this is described by the ${1/N}$ inhomogeneous Landau equation~\citep[see, e.g.\@,][]{Chavanis2013}.
In the present context, it reads
\begin{align}
\frac{\p F (J , t)}{\p t} = \gamma \frac{\p }{\p J} \bigg[ {} & \sum_{k} k \!\! \int \!\! \rd J_{1} \, \big| \Lambda_{k} (\bJ) \big|^{2}
\nonumber
\\
\times {} & \, \deltaD \big( \bk \!\cdot\! \bO \big) \, \bk \!\cdot\! \frac{\p }{\p \bJ} F_{2} (\bJ) \bigg] ,
\label{eq:Landau_1N}
\end{align}
where we dropped the time dependence for clarity.
We refer to Appendix~\ref{app:1_N_prediction}
for the expression of the coupling coefficients, ${ |\Lambda_{k} (\bJ)|^{2} }$,
which are independent of both $N$ and $q$.
In Eq.~\eqref{eq:Landau_1N}, notations are shortened
with the 2-vectors ${ \bJ \!=\! (J , J_{1}) }$,
${ \bO \!=\! (\Omega[J] , \Omega[J_{1}]) }$,
and ${ \bk \!=\! (k , - k) }$,
as well as ${ F_{2} (\bJ) \!=\! F(J) F(J_{1}) }$.
When accounting for collective effects,
Eq.~\eqref{eq:Landau_1N} becomes the inhomogeneous Balescu--Lenard equation~\citep{Heyvaerts2010,Chavanis2012AA}.
The functional form of Eq.~\eqref{eq:Landau_1N} remains the exact same,
and one only has to replace the bare coupling coefficients, ${ |\Lambda_{k} (\bJ)|^{2} }$,
with their (frequency-dependent) dressed analogs, ${ |\Lambdad_{k} (\bJ ; k \Omega(J))|^{2} }$.
In the dynamically hot limit, as expected, one has ${ |\Lambdad_{k}|^{2} \!\to\! |\Lambda_{k}|^{2} }$.
This is detailed in Appendix~\ref{app:LinearResponse},
where we present a new approach to compute ${ |\Lambdad_{k}|^{2} }$.

The Landau equation (Eq.~\ref{eq:Landau_1N}) satisfies
a couple of important conservation laws.
Up to prefactors, these are
\begin{subequations}
\begin{align}
\Gamma {} & = \!\! \int \!\! \rd J \, F (J , t) 
{} & \text{(total circulation)} ;
\label{eq:total_circulation}
\\
L {} & = \!\! \int \!\! \rd J \, J \, F (J , t) 
{} & \text{(total momentum)} ;
\label{eq:total_momentum}
\\
E {} & = \!\! \int \!\! \rd J \, H (J) \, F (J , t) 
{} & \text{(total energy)} .
\label{eq:total_energy}
\end{align}
\label{eq:total_conservation}\end{subequations}
If we define the Boltzmann entropy as
\begin{equation}
S (t) = - \!\! \int \!\! \rd J \, F (J , t) \, \ln \big[ F (J , t) \big] ,
\label{eq:def_entropy}
\end{equation}
the ${1/N}$ Landau equation satisfies an $H$-theorem,
i.e.\ it drives ${ \rd S / \rd t \!\geq\! 0 }$.
Finally, the thermodynamical equilibria,
i.e.\ the Boltzmann DFs, of the form
\begin{equation}
\FB (J) \propto \re^{- \beta H(J) + \gamma J} ,
\label{eq:def_FB}
\end{equation}
are generic steady states of the ${1/N}$ Landau equation,
i.e.\ they satisfy ${ \p \FB / \p t \!=\! 0 }$
when inserted into Eq.~\eqref{eq:Landau_1N}.
There is no guarantee, however,
that the inhomogeneous Landau equation at order ${ 1/N }$
relaxes towards the Boltzmann distribution
because of (local) kinetic blockings, as explained below.

Performing explicitly the sum over $k$ in Eq.~\eqref{eq:Landau_1N},
the ${ 1/N }$ Landau equation can be rewritten as
\begin{align}
\frac{\p F (J , t)}{\p t} = {} & \gamma \frac{\p }{\p J} \bigg[ \!\! \int \!\! \rd J_{1} \, \deltaD \big[ \Omega (J) \!-\! \Omega (J_{1}) \big]
\label{eq:Landau_1N_noK}
\\
\times {} & \, \big| \Lambdatot (J , J_{1}) \big|^{2} \, \bigg( F (J_{1}) \, \frac{\p F}{\p J} \!-\! F (J) \, \frac{\p F}{\p J_{1}} \bigg) \bigg] ,
\nonumber
\end{align}
with
\begin{equation}
\big| \Lambdatot (\bJ) \big|^{2} \!=\! 2 \sum_{k > 0} k \, | \Lambda_{k} (\bJ) |^{2} 
\label{eq:def_Lambdatot}
\end{equation}
the total coupling coefficient.
In that form, Eq.~\eqref{eq:Landau_1N_noK} involves
the simple resonance condition ${ \deltaD [\Omega(J) \!-\! \Omega (J_{1})] }$.

Equation~\eqref{eq:Landau_1N_noK} can also be rewritten
under the form of a Fokker--Planck equation~\citep[see, e.g.\@][]{Chavanis2012AA}, via
\begin{equation}
\frac{\p F (J , t)}{\p t} \!=\! - \frac{\p }{\p J} \bigg[\! D_{1} (J) \, F (J) \!\bigg] \!+\! \frac{1}{2} \frac{\p^{2} }{\p J^{2}} \bigg[\! D_{2} (J) \, \frac{\p F}{\p J} \!\bigg] .
\label{eq:FP_1N}
\end{equation}
Here, $D_{1}$ [resp.\ $D_{2}$] is the first-order [resp.\ second-order]
diffusion coefficient in action space.
In the following, we will focus our interest only on the second-order diffusion coefficient.
It generically reads
\begin{equation}
D_{2} (J) = 2 \gamma \!\! \int \!\! \rd J_{1} \, \deltaD \big[ \Omega (J) \!-\! \Omega (J_{1}) \big] \, \big| \Lambdatot (J , J_{1}) \big|^{2} \, F (J_{1}) .
\label{eq:def_D2}
\end{equation}
Physically, ${ D_{2} (J) }$ describes the diffusion induced by stochastic potential fluctuations.
For example, let us consider a zero-mass test particle,
of initial action $J$, embedded in the present system,.
Then, on long timescales, the growth of the variance of the action increment
for this particle, i.e.\ ${ \Delta J (T) \!=\! J (T) \!-\! J(0) }$,
is given by
\begin{equation}
D_{2} (J) = \lim\limits_{T \to + \infty} \big\langle \big[ \Delta J (T) \big]^{2} \big\rangle / T .
\label{eq:limit_D2}
\end{equation}
Here, the ensemble average is taken over both the realisations
of the background system and the initial phase of the test particle.

Coming back to Eq.~\eqref{eq:Landau_1N_noK}
we note that if the frequency profile is monotonic,
the condition ${ \Omega (J_{1}) \!=\! \Omega(J) }$ implies
${ J_1 \!=\! J }$ and Eq.~\eqref{eq:Landau_1N} vanishes,
i.e.\ ${ \p F / \p t \!=\! 0 }$.
There is no relaxation through ${ 1/N} $ effects.
This is a situation of global kinetic blocking at order ${ 1/N }$.
We note that such a vanishing does not occur
for the diffusion coefficient from Eq.~\eqref{eq:def_D2}.
In the presence of a non-monotonic frequency profile,
for a given $J$, there may exist another ${ J_{1} \!\neq\! J }$
such that ${ \Omega (J_{1}) \!=\! \Omega(J) }$.
In that case, Eq.~\eqref{eq:Landau_1N} does not vanish
and the system can relax through ${1/N}$ effects.
This is the case for profile (1), as visible in Fig.~\ref{fig:DF}.

In that case, assuming that collective effects are negligible,
we find from Eq.~\eqref{eq:Landau_1N_noK} that
the relaxation time,
i.e.\ the time for $F$ to change by order unity,
scales like
\begin{equation}
\Trelax \simeq \Tdyn \, N / q^{2} ,
\label{eq:Trelax_1N}
\end{equation}
with respect to the total number of particles, $N$,
and the system's active fraction, $q$.
As a result, it is useful to introduce the rescaled relaxation rate
and diffusion coefficients
\begin{subequations}
\begin{align}
\mR_{1} (J,t) {} & = \frac{\p N(\!<\!J,t) / \p t}{N / \Tdyn} \, \frac{N}{q^{2}} ,
\label{eq:def_R1}
\\
\mD (J) {} & = \frac{D_{2} (J)}{J_{0}^{2} / \Tdyn} \, \frac{N}{q^{2}} ,
\label{eq:def_mD}
\end{align}
\label{eq:def_rescaled}\end{subequations}
with ${ N(\!<\!J,t) \!=\! (2\pi / \gamma) \int_{0}^{J} \rd \Jp F(\Jp,t) }$
the number of particles with an action smaller than $J$ at time $t$.
Its time derivative is given by ${ \p N(\!<\!J,t) / \p t \!=\! (2 \pi / \gamma) \mF (J,t) }$,
with the Landau flux ${ \p F (J , t) \p t \!=\! \p \mF (J , t) / \p J }$,
following Eq.~\eqref{eq:Landau_1N_noK}.
Here, ${ \mR_{1} (J) }$ and ${ \mD (J) }$ are both dimensionless quantities.
More importantly, the ${1/N}$ Landau equation (Eq.~\ref{eq:Landau_1N_noK}) predicts
that the two of them should be independent of both $N$ and $q$.

\subsection{Linear stability}
\label{sec:1N_Stability}

Let us now ensure that we consider systems that are dynamically hot enough.
As detailed in Appendix~\ref{app:LinearResponse},
the strength of collective effects can be estimated
via the ``dielectric matrix'', ${ \bE_{k} (\omegaR) }$
for ${ \omegaR \!\in\! \mathbb{R} }$.
Then, the dressed coupling coefficients, ${ \Lambdad_{k} }$,
that appear in the ${1/N}$ Balescu--Lenard equation
when compared with the ones from the Landau equation
scale typically like
${ \Lambdad_{k} (\bJ ; \omegaR) \!\sim\! \Lambda_{k} (\bJ) / |\bE_{k} (\omegaR)| }$,
with ${ |\bE_{k} (\omegaR)| \!=\! \det [\bE_{k} (\omegaR)] }$,
the determinant of the dielectric matrix.
In practice, the closer ${ |\bE_{k} (\omegaR)| }$ is from unity,
the weaker are collective effects.
On the contrary, if ${ |\bE_{k} (\omegaR)| \!=\! 0 }$,
then the system supports an infinite amplification,
i.e.\ this is a neutral mode.
Importantly, since it describes collisionless response theory,
we note that ${ |\bE_{k} (\omegaR) | }$ is independent of $N$
and depends only on the active fraction, $q$.

\begin{figure}
\begin{center}
\includegraphics[width=0.45\textwidth]{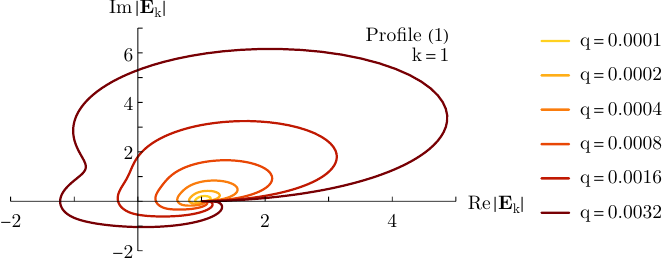}
\caption{Nyquist contour for profile (1) for various values of the active fraction $q$,
for the resonance ${ k \!=\! 1 }$.
We refer to Appendix~\ref{app:LinearResponse} for the definition
of the ``dielectric matrix'', ${ \bE_{k} (\omega) }$.
For ${ q \!\gtrsim\! 0.0010 }$, the system is found to be linearly unstable.
\label{fig:Nyquist_1}}
\end{center}
\end{figure}

In Fig.~\ref{fig:Nyquist_1}, we illustrate the Nyquist contour for profile (1),
i.e.\ the curve of ${ \omegaR \!\mapsto\! |\bE_{k} (\omegaR)| }$ in the complex plane.
In that figure, we observe that for ${ q \!\gtrsim\! 0.0010 }$,
the system is linearly unstable since the Nyquist contour encloses the origin
of the complex plane.\footnote{See, e.g.\@, \cite{Chavanis2012Nyquist}
for a simple illustration of the Nyquist method
in homogeneous systems with long-range interactions.
The possibly unstable profile (1) looks similar to the ``gravity'' case (compare Fig.~\ref{fig:Nyquist_1} with fig.~{2} of~\cite{Chavanis2012Nyquist}),
while the always stable profile (2) looks similar to the ``plasma'' case 
(compare Fig.~\ref{fig:Nyquist_2} with fig.~{21}
of~\cite{Chavanis2012Nyquist}).}
Moreover, we find that for ${ q \!\lesssim\! 0.0003 }$,
the contour remains close to ${ |\bE_{k} (\omegaR)| \!=\! 1 }$,
i.e.\ collective amplification is small.
In all the upcoming explorations, we limit ourselves
to numerical simulations with
${ q \!\leq\! 0.0003 }$,
i.e.\ well within the range of applicability of the ${1/N}$ Landau equation
in which collective effects are neglected.

\subsection{Small-scale contributions}
\label{sec:1N_SmallScale}

Before comparing quantitatively the kinetic prediction
with $N$-body simulations, let us consider the contributions
from small scales in Eq.~\eqref{eq:Landau_1N}.
Indeed, that equation involves an infinite sum over the resonance numbers $k$,
so that one should ensure that the limit ${ k \!\to\! + \infty }$ is meaningful.

\begin{figure}
\begin{center}
\includegraphics[width=0.45\textwidth]{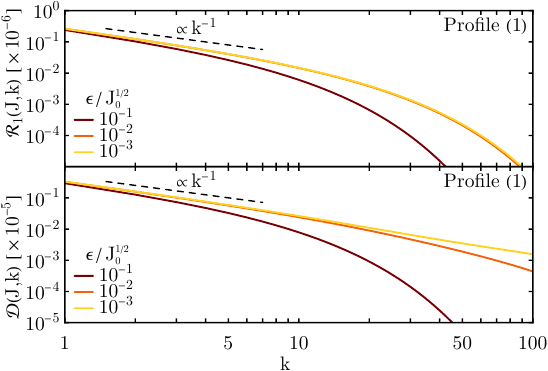}
\caption{Contribution to the relaxation rate, ${ \mR_{1} (J , k) }$ (Eq.~\ref{eq:def_R1k}, top panel),
and diffusion coefficient, ${ \mD (J , k) }$ (Eq.~\ref{eq:def_mDk}, bottom panel), 
evaluated in ${ J / J_{0} \!=\! 1.1 }$,
as a function of the resonance number, $k$.
The different colours correspond to different values of the softening length, $\eps$.
For small $k$, we recover the expected scaling in $k^{-1}$,
before it being damped by softening.
\label{fig:Coulomb_1}}
\end{center}
\end{figure}

Following Eq.~\eqref{eq:def_Lambdatot},
we can naturally decompose the relaxation rate
and diffusion coefficient from Eq.~\eqref{eq:def_rescaled} as
\begin{subequations}
\begin{align}
\mR_{1} (J) {} & = \sum_{k > 0} \mR_{1} (J, k) ,
\label{eq:def_R1k}
\\
\mD (J) {} & = \sum_{k > 0} \mD (J , k) ,
\label{eq:def_mDk}
\end{align}
\label{eq:def_with_k}\end{subequations}
where we dropped the dependence with respect to $t$ to shorten the notations.
In Fig.~\ref{fig:Coulomb_1},
we illustrate the dependence with $k$
of ${ \mR_{1} (J , k) }$ and ${ \mD (J , k) }$,
when evaluated in ${ J \!=\! 1.1 \, J_{0} }$.
We emphasise that the considered action satisfies a non-local resonance condition (Fig.~\ref{fig:DF}).

From Fig.~\ref{fig:Coulomb_1}, we can make three observations.
(i) For $k$ small enough, one finds ${ \mR_{1} (J, k) , \mD (J, k) \!\propto\! k^{-1} }$,
in agreement with the decay rate
of the coupling coefficient, ${ k \!\mapsto\! U_{k} (J , J_{1}) }$,
that can directly be inferred from Eq.~\eqref{eq:Uk_generic},
as detailed in Appendix~\ref{app:1_N_prediction}.
(ii) As $k$ increases, we find that ${ k \!\mapsto\! \mR_{1} (J, k) }$ decreases
faster than $k^{-1}$. This comes from the fact that for $k$ large enough,
the difference between the two actions, $J$ and $J_{1}$, that are in resonance in Eq.~\eqref{eq:Landau_1N_noK}
leads to a decay faster than algebraic (see Eq.~\ref{eq:Uk_generic}).
Indeed, following Eq.~\eqref{eq:Uk_generic},
and assuming ${ J_{1} \!\leq\! J }$,
we find ${ U_{k} (J,J_{1}) \!\propto\! (J_{1} / J)^{k} / |k| }$,
hence driving an exponentially fast decay.
(iii) On the contrary, as $k$ increases,
${ k \!\mapsto\! \mD (J,k) }$
continues to scale like $k^{-1}$.
This is because local resonances, i.e.\ ${ J_{1} \!=\! J }$ in Eq.~\eqref{eq:def_D2}, always contribute to the diffusion coefficient.
Indeed, following Eq.~\eqref{eq:Uk_generic}, for local resonances,
one finds ${ U_{k} (J , J) \!\propto\! 1/|k| }$.
(iv) Finally, as one increases the softening length, $\eps$
(Eq.~\ref{eq:def_U_softened}), the decay rate with respect to $k$
gets faster, for both the relaxation rate and the diffusion coefficient.
Indeed, even when evaluated for local resonances,
Eq.~\eqref{eq:Uk_soft} roughly gives the scaling
${ U_{k} (J,J) \!\propto\! (1 \!-\! \eps / J)^{k} / |k| }$,
hence imposing an exponentially fast decay.
This makes perfect sense physically,
since softening naturally washes out contributions from small scales.

Overall, in the present regime,
the ${1/N}$ Landau diffusion coefficient without softening
suffers from a small-scale logarithmic divergence.
This is similar to the Coulomb logarithm
that plagues the ${1/N}$ Landau equation when applied
to ${3D}$ self-gravitating systems~\citep[see, e.g.\@][]{Chavanis2013}.
In the presence of softening,
this small-scale divergence of ${ \mD (J,k) }$ is naturally washed out.
By contrast, the ${1/N}$ Landau flux, because it is unaffected by local resonances,
does not suffer from any small-scale divergence, even without softening.
We also refer to Appendix~\ref{app:reglog} for a discussion of
how resonance broadening also cures this small-scale divergence.
In practice, in all the upcoming simulations,
we will consider dynamics driven by the softened interaction potential
from Eq.~\eqref{eq:def_U_softened},
fixing the softening length to ${ \eps \!=\! 10^{-2} J_{0} }$.

\subsection{Relaxation and diffusion}
\label{sec:1N_RelaxationRate}

\begin{figure}
\begin{center}
\includegraphics[width=0.45\textwidth]{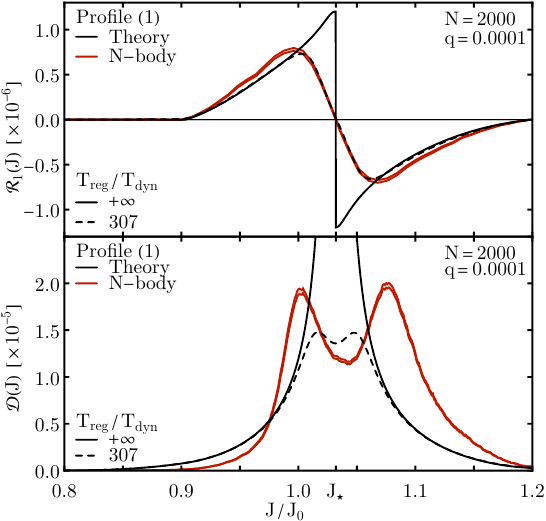}
\caption{Initial relaxation rate in action space, ${ \mR_{1} (J) }$ (Eq.~\ref{eq:def_R1}, top panel),
and diffusion coefficient, ${ \mD (J) }$ (Eq.~\ref{eq:def_mD}, bottom panel),
for profile (1).
For the simulations, the uncertainty is illustrated with the 16\% and 84\% contours
obtained by bootstrap over the available realisations.
We refer to Appendix~\ref{app:NumericalSimulations} for details on the numerical setup.
For the prediction, the full line corresponds to the ${ 1/N }$ Landau equation
(Eqs.~\ref{eq:Landau_1N} and~\ref{eq:def_D2}),
while the dashed line is from the ``regularised'' ${ 1/N }$ Landau equation
(Eqs.~\ref{eq:Landau_1N_REG} and~\ref{eq:D2_1N_REG}),
with the regularised time, ${ \Treg / \Tdyn \!=\! 307 }$,
adjusted to the slope around the sign switch of ${ \mR_{1} (J) }$.
Close to the extremum of the frequency profile, broadening effects
play a crucial role.
\label{fig:Landau_1}}
\end{center}
\end{figure}

We are now set to compare the prediction from Eq.~\eqref{eq:Landau_1N}
with measurements in $N$-body simulations.
This is presented in Fig.~\ref{fig:Landau_1},
where we investigate both the relaxation rate and the diffusion coefficient.
A couple of remarks can be made on Fig.~\ref{fig:Landau_1}.

\subsubsection{Analytical predictions}
\label{sec:1N_RelaxationRateAnalytical}

For now, let us focus solely on the kinetic predictions,
for the Landau flux (Eq.~\ref{eq:Landau_1N}) in the top panel
and the diffusion coefficient (Eq.~\ref{eq:def_D2}) in the bottom panel.
First, for ${ J / J_{0} \!\lesssim\! 0.9 }$, it predicts ${ \mR_{1} (J) \!=\! 0 }$,
while, for the same range of action, one finds ${ \mD (J) \!>\! 0 }$.
Glancing back at Fig.~\ref{fig:DF},
this corresponds to a range of actions, $J$,
for which the resonance condition, ${ \Omega (J_{1}) \!=\! \Omega (J) }$ (Eq.~\ref{eq:Landau_1N_noK}),
imposes ${ J_{1} \!=\! J }$.
As such, only local resonances are permitted by the mean frequency profile.
These local resonances lead to the local vanishing of Eq.~\eqref{eq:Landau_1N_noK}
because of the antisymmetry of the collision kernel.
On the contrary, such local resonances always contribute to the diffusion coefficient.
As a result, the diffusion coefficient is non-zero over the whole action domain considered.
At order ${1/N}$, because its overall relaxation rate vanishes,
the action region, ${ J / J_{0} \!\lesssim\! 0.9 }$ does not relax towards the Boltzmann distribution. This is a local kinetic blocking.
As such, to relax towards the statistical equilibrium state, $1/N^2$ effects must be taken into account.
This will be studied in Section~\ref{sec:1_N2_dynamics}
for a purely monotonic frequency profile.

Then, in Fig.~\ref{fig:Landau_1},
for ${ J \!\to\! \Jstar \!\simeq\! 1.03 \, J_{0} }$,
we find that the kinetic prediction for ${ \mR_{1} (J) }$ abruptly changes sign,
while the prediction for ${ \mD (J) }$ diverges locally.
This is due to the extremum of the frequency profile
in ${ J \!=\! \Jstar }$, as visible in Fig.~\ref{fig:DF}.
Let us now detail the origin of such a behaviour.

For ${ J \!\to\! \Jstar }$,
we may, up to a constant, approximate the frequency profile
with the quadratic dependence
${ \Omega (J) \!\simeq\! (J \!-\! \Jstar)^{2} }$, see Fig.~\ref{fig:DF}.
As a result, when evaluated in ${ J \!=\! \Jstar \!+\! \veps }$,
Eqs.~\eqref{eq:Landau_1N_noK} and~\eqref{eq:def_D2}
predict for the relaxation rate and diffusion coefficient a behaviour of the form
\begin{subequations}
\begin{align}
\mR_{1} (\Jstar \!+\! \veps) {} & \propto \!\! \int \!\! \rd J_{1} \, \deltaD \big[ \veps^{2} \!-\! \big( J_{1} \!-\! \Jstar \big)^{2} \big]
\label{eq:DL_flux_mR}
\\
\times {} & \big[ F' (\Jstar \!+\! \veps) F (J_{1}) \!-\! F (\Jstar \!+\! \veps) F' (J_{1}) \big] ,
\nonumber
\\
\mD (\Jstar \!+\! \veps) {} & \propto \!\! \int \!\! \rd J_{1} \, \deltaD \big[ \veps^{2} \!-\! \big( J_{1} \!-\! \Jstar \big)^{2} \big] \, F (J_{1}) .
\label{eq:DL_flux_mD}
\end{align}
\label{eq:DL_flux}\end{subequations}
where we used the notation ${ F' (J) \!=\! \p F / \p J }$,
and dropped the coupling coefficient, ${ |\Lambdatot|^{2} }$,
taking it to be constant.

To perform the integral over action space,
we use the generic relation
\begin{equation}
\deltaD [f(J)] = \sum_{\Jr} \deltaD [J \!-\! \Jr] / |\Omega' (\Jr)| ,
\label{eq:solve_resonance_condition}
\end{equation}
where the sum runs over the resonant actions, $\Jr$, such that ${ f (\Jr) \!=\! 0 }$.
In the present case, the resonance condition from Eq.~\eqref{eq:DL_flux}
imposes ${ J_{1} \!=\! \Jstar \!\pm\! \veps }$.
As previously argued, we find that the contribution from the local resonance
${ J_{1} \!=\! \Jstar \!+\! \veps \!=\! J }$ does not contribute to the flux, ${ \mR_{1} (J) }$,
while it contributes to the diffusion coefficient, ${ \mD (J) }$.
The non-local resonance ${ J_{1} \!=\! \Jstar \!-\! \veps }$ contributes to both quantities.
Equation~\eqref{eq:DL_flux} becomes
\begin{subequations}
\begin{align}
\mR_{1} (\Jstar \!+\! \veps) {} & \propto \frac{1}{|\veps|} \, \big[ F' (\Jstar \!+\! \veps) F (\Jstar \!-\! \veps) \!-\! F (\Jstar \!+\! \veps) F' (\Jstar \!-\! \veps) \big] ,
\label{eq:DL_flux_continued_mR}
\\
\mD (\Jstar \!+\! \veps) {} & \propto \frac{1}{|\veps|} \, \big[ F (\Jstar \!+\! \veps) + F (\Jstar \!-\! \veps) \big] .
\label{eq:DL_flux_continued_mD}
\end{align}
\label{eq:DL_flux_continued}\end{subequations}
In the limit ${ \veps \!\to\! 0 }$, i.e.\ as one nears the extremum
of the frequency profile, one finds the asymptotic behaviours
\begin{subequations}
\begin{align}
\mR_{1} (\Jstar \!+\! \veps) {} & \simeq \Sign[\veps]
& \!\!\!\!\!\!\!\! \text{for} \quad
\veps \to 0 ,
\label{eq:DL_flux_final_mR}
\\
\mD (\Jstar \!+\! \veps) {} & \simeq \frac{1}{|\veps|}
& \!\!\!\!\!\!\!\! \text{for} \quad
\veps \to 0 .
\label{eq:DL_flux_final_mD}
\end{align}
\label{eq:DL_flux_final}\end{subequations}
As such, as visible in Fig.~\eqref{fig:Landau_1},
the ${1/N}$ Landau equation predicts that
(i) the flux in action space is discontinuous
near the extremum of the frequency profile,
but suffers from no divergence there;
(ii) the diffusion coefficient in action space
suffers from a genuine local divergence.

\subsubsection{Numerical measurements}
\label{sec:1N_RelaxationRateNumerical}

We can now compare the predictions
of the flux and diffusion coefficients
with measurements in $N$-body simulations.
First, for the flux (top panel of Fig.~\ref{fig:Landau_1}),
we recover that for ${ J / J_{0} \!\lesssim\! 0.9 }$,
relaxation is greatly delayed.
This is the imprint of the local kinetic blocking.
Second, as one considers larger action,
say ${ 0.9 \!\lesssim\! J / J_{0} \!\lesssim\! 1.0 }$,
we find a good agreement between the numerical simulations
and the kinetic prediction,
for both the flux and the diffusion coefficient.
However, as ${ J }$ nears $\Jstar$,
the location of the extremum of the frequency profile,
we find that the kinetic prediction overestimates the numerical measurements.
Indeed, the simulations do not exhibit
the discontinuity of the flux
nor the divergence of the diffusion coefficient,
predicted by the ${1/N}$ Landau equation.
In particular, the numerically-measured flux, ${ \mR_{1} (J) }$,
exhibits a rather smooth change of sign,
while the numerically-measured diffusion coefficient, ${ \mD (J) }$,
presents a two-hump shape that only reaches some finite amplitude.
Finally, further out in action space,
i.e.\ for ${ J / J_{0} \!\gtrsim\! 1.1 }$,
we find that the kinetic prediction for the flux
matches with the $N$-body simulations,
while for the diffusion coefficient, some mismatch between the two remains.
We note that this domain corresponds to actions, $J$,
whose orbital frequency, ${ \Omega (J) }$,
is still close to the extremum one, ${ \Omega (\Jstar) }$.

The mismatch observed in Fig.~\ref{fig:Landau_1}
is to be attributed to the limitations of some of the hypotheses
made in the derivation of the ${1/N}$ Landau equation.
In Fig.~\ref{fig:Landau_1}, we present a prediction with a ``regularised''
Landau equation.
It offers a much more satisfactory agreement
with the numerical simulations.
For the diffusion coefficient, the match is improved, though far from ideal.
Nonetheless, this heuristic theory still manages
to reproduce the smooth change of sign of the flux
and the two-hump structure of the diffusion coefficient.
The exploration of such ``regularised'' kinetic predictions
is the topic of Section~\ref{sec:1N_ResonanceBroadening}.

\subsubsection{Comparison with other ${1D}$ systems}
\label{sec:1N_RelaxationRateOther}

Naturally, one could wonder which specific
features of the current setup make it more akin
to such discontinuities and the need for regularisation.
Indeed, it was shown for ${1D}$ gravity~\citep{Roule+2022}
and the \HMF\ model~\citep{Benetti+2017}
that the ${1/N}$ Landau equation
can provide an excellent description of ${1D}$ relaxation processes.
We briefly review here some of the key differences
between these systems.

In the case of ${ 1D }$ gravity,
there is no divergence on small-scale of the kinetic prediction~\citep[see, e.g.\@, eq.~{(A7)} in][]{Roule+2022}.
In addition, the frequency profile 
of a typical stable quasistationary state exhibits no extremum~\citep[see fig.~{1} in][]{Roule+2022}.
Physically, this means that the contribution from phase mixing never vanishes,
so that there is always a non-zero shear in action space.
For these two reasons,
the ${1/N}$ Landau equation
produces no local discontinuities or divergences in action space.
For that system, one finds then an excellent match
between the ${1/N}$ kinetic prediction and numerical measurements,
both for the diffusion coefficient~\citep[see fig.~{2} in][]{Roule+2022}
and the relaxation rate~\citep[see fig.~{3} in][]{Roule+2022}.

Let us now consider the case of the \HMF\ model~\citep{Antoni+1995}.
Since the pairwise interaction in that system
involves only the ${ k \!=\! \pm 1 }$ large-scale harmonics,
the ${1/N}$ Landau equation does not suffer from any divergence on small scales.
In the limit of a homogeneous quasistationary state, i.e.\ ${ F \!=\! F(v) }$,
the system's action is ${ J \!\sim\! v }$,
and the frequency profile is simply ${ \Omega (J) \!\sim\! \p_{v} [\tfrac{1}{2} v^{2}] \!\sim\! v }$,
making it strictly monotonic.
In the homogeneous limit, the \HMF\ model is therefore submitted
to a global kinetic blocking of its ${1/N}$ dynamics,
similarly to profile (2) here.
In that case, the overall relaxation is driven only by ${1/N^{2}}$ effects~\citep{Rocha+2014}.
In the limit of weak collective amplification,
the ${1/N^{2}}$ Landau equation offers a convincing match
with the numerical measurements~\citep[see fig.~{2} in][]{Fouvry+2019}.

Finally, let us consider the case of the \HMF\ model
when in an inhomogeneous,
i.e.\ magnetised, quasistationary state.
Here again, there is no divergence on small scales of the ${1/N}$ kinetic theory.
The typical frequency profile of such a system exhibits
two striking features~\citep[see, e.g.\@, fig.~{3} in][]{Benetti+2017}:
(i) For ${ J \!\to\! 0 }$, i.e.\ in the centre of the trapped region,
one has ${ \Omega' (J) \!\to\! 0 }$;
(ii) For ${ J \!\to\! \Jsep }$, i.e.\ at the system's separatrix,
one has ${ |\Omega'(J)| \!\to\! + \infty }$.

Naively, one could fear that the vanishing of the frequency gradient
in the system's centre would lead
to the same asymptotic behaviour as in Eq.~\eqref{eq:DL_flux_final}.
In practice, for the inhomogeneous \HMF\ model,
one finds that the pairwise coupling coefficient satisfies
${ \lim_{J \to 0} U_{k} (J ,J) \!=\! 0 }$~\citep[see eqs.~{(45)} and {(46)} in][]{Benetti+2017}.
Physically, this makes sense since orbits in the system's centre
correspond to particles that are exactly at rest.
As a result, the Fourier transform in angle
of the interaction potential between two such orbits
therefore vanishes~\citep[see eq. {(44)} in][]{Benetti+2017}.
This differs greatly from the present case of point vortices
where, at the extremum of the frequency profile,
one finds that ${ \lim_{J \to \Jstar} U_{k} (J , J) }$ is non-zero.
Because of the vanishing of the pairwise coupling coefficient
in the centre of the inhomogeneous \HMF\ model,
one finds that both the relaxation rate and the diffusion coefficient vanish there.
Therefore, the associated ${1/N}$ kinetic prediction
exhibits no discontinuities nor divergences,
and matches with numerical measurements~\citep[see, e.g.\@, fig.~{6} in][]{Benetti+2017}.

At the location of the \HMF\ separatrix,
we note that the gradient of the frequency profile
tends to infinity.
Physically, this means that close the separatrix,
pairwise resonances are swept extremely fast,
rather than extremely slow as is the case close to a vanishing frequency gradient.
As such, the type of discontinuities and divergences
emphasised in Eq.~\eqref{eq:DL_flux_final} are not to be expected
close to the \HMF\ separatrix.
Nonetheless, we note that this rapidly varying frequency profile
leads to localised, but finite, ``spikes''
in the prediction for the Landau diffusion coefficient.
These are clearly visible in fig.~{2} of~\cite{Fouvry+2018}
and are in agreement there with $N$-body measurements.
Finally, we point out that the non-monotonicity
of the frequency profile of the inhomogeneous \HMF\ model
is essential to prevent this system from suffering from a global kinetic blocking.
Indeed, it is the non-local resonant couplings
between particles located on each side of the separatrix
that are solely responsible for driving the overall ${1/N}$ relaxation.

\subsection{Resonance broadening}
\label{sec:1N_ResonanceBroadening}

The ${1/N}$ Landau equation (Eq.~\ref{eq:Landau_1N_noK})
involves the sharp resonance condition,
${ \deltaD [ \Omega (J) \!-\! \Omega (J_{1}) ] }$.
Mathematically speaking, this sharp resonant Dirac comes
from the fact that the derivation of the Landau equation
requires the propagation of the dynamics
of the \DF\@'s fluctuations, at linear order,
up to infinite dynamical times~\citep[see, e.g.\@,][]{Chavanis2012AA}.
Yet, as a result of nonlinearities and diffusion, such an assumption is only partially true.
This effect can be responsible in particular for ``resonance broadening''~\citep[see, e.g.\@][]{Weinstock1969},
as we will now explore, in a heuristic fashion.

\subsubsection{Heuristic regularisation of the Landau equation}
\label{sec:1N_RegLandau}

Typically, the derivation of the Landau equation starts with the introduction
of the system's empirical DF, ${ \Fd \!=\! \sum_{i} \gamma_{i} \deltaD [\br \!-\! \br_{i} (t)] }$.
This DF evolves according to the Klimontovich equation, ${ \p_{t} \Fd \!+\! [\Fd , \Hd] \!=\! 0 }$,
with ${ \Hd \!=\! \Hd [\Fd] }$ the system's (specific) empirical Hamiltonian,
and ${ [\cdot , \cdot] }$ the Poisson bracket~\citep{Chavanis2012AA}.
The empirical DF is then expanded into
${ \Fd \!=\! F \!+\! \delta F }$, with $F$ the (smooth) ensemble-averaged DF from Section~\ref{sec:InitialConditions}, and ${ \delta F }$ its fluctuations.
The crucial step of the quasilinear assumption is to assume
that the dynamics of ${ \delta F }$ can be expanded at linear order,
to become
\begin{equation}
\frac{\p \delta F}{\p t} + [\delta F , H] + [F , \delta H] = 0 ,
\label{eq:linear_deltaF}
\end{equation}
with $H$ the system's mean Hamiltonian,
and ${ \delta H \!=\! \delta H [\delta F] }$
the fluctuations in the Hamiltonian generated by ${ \delta F }$.
When considered within angle-action coordinates,
Eq.~\eqref{eq:linear_deltaF} is straightforward to solve
and gives a solution of the form
\begin{equation}
\delta F_{k} (J , t) \propto \delta F_{k} (J , 0) \, \re^{- \ri k \Omega (J) t} ,
\label{eq:shape_deltaF_time}
\end{equation}
with ${ \delta F_{k} }$ the Fourier transform
in angle of ${ \delta F }$,
for the harmonic $k$.
Physically, this solution assumes that the DF fluctuations
follow the system's unperturbed mean field orbits.
Finally, by sending the dynamical time to infinity,
we find that
\begin{equation}
\mathrm{Re} \bigg[ \lim\limits_{T \to + \infty} \!\! \int_{0}^{T} \!\! \rd t \, \re^{- \ri \omegaR t} \bigg] \simeq \pi \, \deltaD (\omegaR) ,
\label{eq:t_to_infinity}
\end{equation}
with ${ \omegaR \!=\! \Omega (J) \!-\! \Omega (J_{1}) }$
the resonance frequency.
Therefore, in Eq.~\eqref{eq:t_to_infinity},
we recover, heuristically, the sharp resonance condition from Eq.~\eqref{eq:Landau_1N_noK}.
To sum up, the ${1/N}$ Landau equation involves a Dirac delta
as a result of two assumptions (i) the DF fluctuations are propagated linearly,
i.e.\ along the unperturbed orbits; (ii) and this holds for infinitely long dynamical times.

Yet, these two assumptions cannot hold for so long,
since we neglected the contributions from the nonlinear term, ${ [\delta F , \delta H] }$,
in Eq.~\eqref{eq:linear_deltaF}. Heuristically, let us therefore assume
that Eq.~\eqref{eq:shape_deltaF_time} must be weighted down
by an exponential damping factor on some ``regularisation'' timescale, $\Treg$.
The dynamics of the DF fluctuations would become
\begin{equation}
\delta F_{k} (J , t) \!\propto\! \delta F_{k} (J , 0) \, \re^{- \ri k \Omega (J) t} \, \re^{- k t / \Treg} .
\label{eq:shape_deltaF_exp}
\end{equation}
Physically, this means that after a time ${ \Treg / k }$,
fluctuations on scale $k$ are scattered away from their mean field orbit.
This prevents the further build up of resonant interactions.
Performing the same integral as in Eq.~\eqref{eq:t_to_infinity}, we obtain
\begin{equation}
\mathrm{Re} \bigg[ \lim\limits_{T \to + \infty} \!\! \int_{0}^{T} \!\! \rd t \, \re^{- \ri \omegaR t} \, \re^{- t / \Treg} \bigg] \simeq \pi \, \delta_{\Treg} (\omegaR) ,
\label{eq:t_to_infinity_nl}
\end{equation}
where we introduced the Lorentzian
\begin{equation}
\delta_{\Treg} (\omega) = \frac{1}{\pi} \frac{\Treg}{1 + (\omega \Treg)^{2}}.
\label{eq:def_Lorentzian}
\end{equation}
Naturally, in the limit ${ \Treg \!\to\! + \infty }$,
one has ${ \delta_{\Treg} (\omega) \!\to\! \deltaD (\omega) }$,
hence recovering a sharp resonance condition.
We refer to Appendix~\ref{app:Landau_from_stochastic}
for an alternative approach to justify
the appearance of the broadened Dirac from Eq.~\eqref{eq:def_Lorentzian} by taking into account the stochasticity of the particles trajectories.

Armed with this regularised resonance condition,
we can then, heuristically, introduce ``resonance broadening''
in Eq.~\eqref{eq:Landau_1N_noK}
by replacing its Dirac delta with this broadened condition.
It becomes
\begin{align}
\frac{\p F (J , t)}{\p t} = {} & \gamma \frac{\p }{\p J} \bigg[ \!\! \int \!\! \rd J_{1} \, \delta_{\Treg} \big[ \Omega (J) \!-\! \Omega (J_{1}) \big]
\label{eq:Landau_1N_REG}
\\
\times {} & \, \big| \Lambdatot (J , J_{1}) \big|^{2} \, \bigg( F (J_{1}) \, \frac{\p F}{\p J} \!-\! F (J) \, \frac{\p F}{\p J_{1}} \bigg) \bigg].
\nonumber
\end{align}
Similarly, following the same line of thought,
the diffusion coefficient from Eq.~\eqref{eq:def_D2}
can be regularised into
\begin{equation}
D_{2} (J) = 2 \gamma \!\! \int \!\! \rd J_{1} \, \delta_{\Treg} \big[ \Omega (J) \!-\! \Omega (J_{1}) \big] \big| \Lambdatot (J , J_{1}) \big|^{2} \, F (J_{1}) .
\label{eq:D2_1N_REG}
\end{equation}
As a side note, we point out that the broadened Dirac in Eq.~\eqref{eq:def_Lorentzian}
is always positive.
This ensures that the regularised diffusion coefficient
from Eq.~\eqref{eq:D2_1N_REG} is always positive.
This is an important requirement
to comply with the physical definition of the diffusion coefficient from Eq.~\eqref{eq:limit_D2}.
Of course, we stress that Eqs.~\eqref{eq:Landau_1N_REG} and~\eqref{eq:D2_1N_REG}
are only heuristic expressions.
Surely, they would deserve further theoretical developments.
For example, Eq.~\eqref{eq:Landau_1N_REG} does not conserve the total energy,
contrary to Eq.~\eqref{eq:Landau_1N_noK}.
Some additional properties of the regularised Landau equation (Eq.~\ref{eq:Landau_1N_REG})
are also discussed in Appendix~\ref{app:reglog},
in particular how it regularises both the generic divergence of the diffusion coefficient
on small scales (Appendix~\ref{app:reg_div}),
and the divergence of the diffusion coefficient at the extremum of the frequency profile (Appendix~\ref{app:reg_div_Jstar}).

\begin{figure}
\begin{center}
\includegraphics[width=0.45\textwidth]{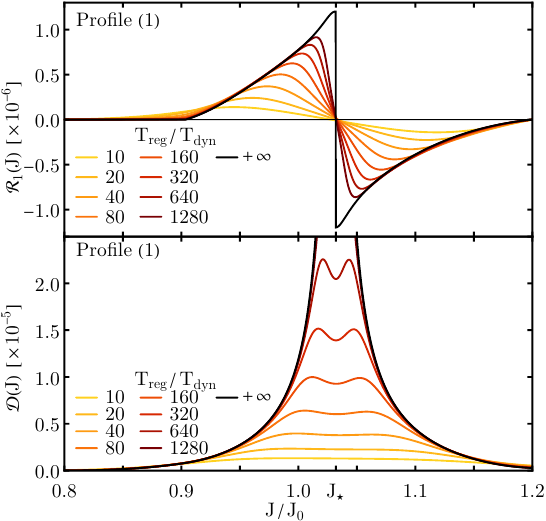}
\caption{Predicted initial regularised relaxation rate (Eq.~\ref{eq:Landau_1N_REG}, top panel)
and diffusion coefficient (Eq.~\ref{eq:D2_1N_REG}, bottom panel),
for profile (1),
as one varies the regularisation time $\Treg$,
using the same convention as in Fig.~\ref{fig:Landau_1}.
For ${ \Treg \!\to\! +\infty }$,
one recovers the ${1/N}$ Landau prediction from Eqs.~\eqref{eq:Landau_1N_noK}
and~\eqref{eq:def_D2}.
\label{fig:Flux_Treg_1}}
\end{center}
\end{figure}

In Fig.~\ref{fig:Flux_Treg_1},
we compute the regularised relaxation rate and diffusion coefficients
predicted by Eqs.~\eqref{eq:Landau_1N_REG} and~\eqref{eq:D2_1N_REG},
as one varies the regularisation timescale, $\Treg$.
In that figure, we find that broadening effects smooth out
the discontinuity and local divergence
predicted by the ${1/N}$ Landau equation (Eq.~\ref{eq:Landau_1N_noK})
close to the extremum of the frequency profile.
Reassuringly, we also find that as ${ \Treg \!\to\! + \infty }$,
the regularised prediction converges to the (sharp) Landau prediction.
As such, varying $\Treg$ in Eq.~\eqref{eq:Landau_1N_REG}
allows for a smooth transition between
the usual ${1/N}$ Landau equation and the nonlinear regime of resonance broadening. In Appendices~\ref{app:reglog} and~\ref{app:t}, we discuss this connexion further.
In particular, we link our result with the ones obtained in previous literature
in the regime of sheared and unsheared flows.

\subsubsection{Finite-time diffusion coefficients}
\label{sec:1N_FiniteTimeDiff}

In order to visualise the appearance of resonance broadening
in the simulations, taking inspiration from Eq.~\eqref{eq:limit_D2},
let us introduce the finite-time diffusion coefficient as
\begin{equation}
D_{2} (J , T) = \frac{\big\langle \big[ \Delta J (T) \big]^{2} \big\rangle}{T} .
\label{eq:def_DT}
\end{equation}
In practice, the ${1/N}$ Landau equation predicts
that ${ \lim_{T \to +\infty} D_{2} (J , T) \!=\! D_{2} (J) }$,
with the Landau diffusion coefficient given by Eq.~\eqref{eq:def_D2}.
Deviations from this limit are therefore to be attributed
to nonlinear contributions spoiling some of the underlying
assumptions of the ${1/N}$ Landau equation.
Naturally, from ${ D_{2} (J , T) }$, one can follow Eq.~\eqref{eq:def_mD}
and readily define the associated rescaled finite-time diffusion coefficient,
${ \mD (J , T) }$.

\begin{figure}
\begin{center}
\includegraphics[width=0.45\textwidth]{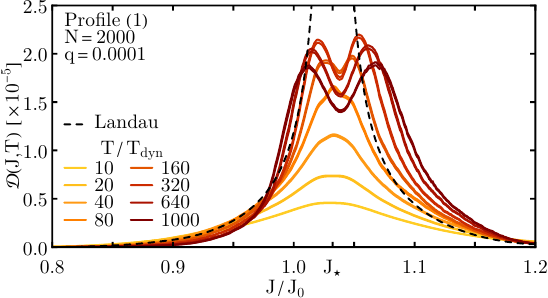}
\caption{Illustration of the finite-time diffusion coefficient,
${ \mD (J , T) }$ (Eq.~\ref{eq:def_DT}),
as measured in $N$-body simulations
for various finite times $T$, for profile (1).
The dashed line is the ${1/N}$ Landau prediction from Eq.~\eqref{eq:def_D2}.
We refer to Appendix~\ref{app:NumericalSimulations} for details on the numerical setup.
As $T$ increases, the finite-time diffusion coefficient grows,
but eventually saturates to some finite amplitude,
as a result of resonance broadening.
\label{fig:DeltaJSQ_T}}
\end{center}
\end{figure}

In Fig.~\ref{fig:DeltaJSQ_T}, we present numerical measurements
of the finite-time diffusion coefficients, ${ \mD (J , T) }$,
for various times $T$.
In that figure, we note that, as one increases $T$,
the finite-time diffusion coefficients start by following
more and more closely the ${1/N}$ Landau diffusion coefficient.
Yet, for ${ T / \Tdyn \!\gtrsim\! 320 }$,
i.e.\ a timescale similar to the value of $\Treg$ introduced in Fig.~\ref{fig:Landau_1},
we note that
(i) the growth of ${ \mD (J , T) }$ halts;
(ii) it starts to exhibit a two-hump structure
reminiscent of the one of the regularised prediction from Fig.~\ref{fig:Flux_Treg_1}.
As one pushes $T$ further, the numerically-measured finite-time diffusion coefficients
even start to decay. This is because for such long time separations,
the test particles have now entered their ``diffusive'' regime,
i.e.\ ${ T \!\to\! \mD (J , T) }$ growing linearly in time,
as illustrated numerically in Fig.~\ref{fig:DeltaJSQ_time_1}
of Appendix~\ref{app:NumericalSimulations}.
In addition, we note that the mismatch between the finite-time diffusion coefficient
and the ${1/N}$ Landau prediction is the strongest
for ${ J \!\geq\! \Jstar }$ rather than for ${ J \!\leq\! \Jstar }$.
This is because particles with ${ J \!\geq\! \Jstar }$
are the ones with an orbital frequency, ${ \Omega (J) }$,
closer to the extremum frequency, ${ \Omega (\Jstar) }$,
as visible in Fig.~\ref{fig:DF}.
Overall, we argue that two-hump structure in Fig.~\ref{fig:DeltaJSQ_T}
along with the saturation to finite amplitude
are numerical testimonies that resonance broadening effects
close to the extremum of the frequency profile
play an instrumental role in the relaxation of this system.

\subsubsection{Numerical comparison}
\label{sec:1N_NumericalExploration}

Armed with a regularised kinetic equation (Eq.~\ref{eq:Landau_1N_REG}),
we may use it to compare with $N$-body simulations.
This is first presented in Fig.~\ref{fig:Landau_1},
where we adjusted the considered $\Treg$
to match the slope of the numerical simulations,
near the frequency extremum (Appendix~\ref{app:1_N_prediction}).
Interestingly, we find that Eq.~\eqref{eq:Landau_1N_REG},
although heuristic, offers a satisfactory match with the numerical measurements,
over the whole action domain.
This highlights the fact that nonlinear broadening effects
are likely responsible for the effective smoothing of the relaxation rate
near the extremum of the orbital frequency.
We point out that, here, $\Treg$ was adjusted only to match
the central slope of the relaxation rate in the top panel of Fig.~\ref{fig:Landau_1}.
We then used the exact same regularisation time
in the bottom panel of Fig.~\ref{fig:Landau_1}
for the diffusion coefficient.
There, we find that the agreement between the numerical measurement
and the heuristic prediction is less convincing.
Obtaining a better match requires the development
of a much more involved kinetic prediction,
possibly through renormalisation techniques~\citep[see, e.g.\@,][and references therein]{Krommes2002}.
This will be the topic of future work.

\begin{figure}
\begin{center}
\includegraphics[width=0.49\textwidth]{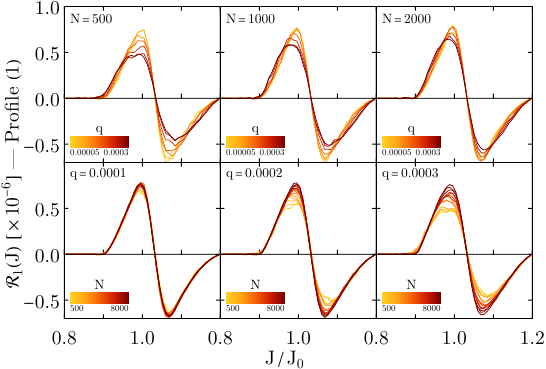}
\caption{Relaxation rate for profile (1),
using the same convention as in Fig.~\ref{fig:Landau_1},
as one varies $q$ for fixed $N$ (top),
or varies $N$ for fixed $q$ (bottom).
Here, we represent only the median value over the available realisations.
As one increases $N$ and decreases $q$,
the relaxation rate increases, but only up to some saturation.
\label{fig:Flux_Grid_1}}
\end{center}
\end{figure}

In practice, in Fig.~\ref{fig:Landau_1},
we presented a single value of ${ (N,q) }$.
Let us now explore the dependence of this broadening
as one varies the total number of particles, $N$,
and the system's active fraction, $q$.
In Fig.~\ref{fig:Flux_Grid_1},
we explore the dependence of the relaxation rate, ${ \mR_{1} (J) }$,
as one varies $N$ and $q$.
First, in that figure, following the rescaling from Eq.~\eqref{eq:def_R1},
we find that the dependence with respect to ${ (N,q) }$ is only very mild.
This emphasises that, indeed, ${1/N}$ effects
are responsible for this system's relaxation.
Yet, as one increases $N$ and/or lowers $q$,
one can tentatively note some increase in the overall amplitude
of the measured relaxation rate.
However, this increase is only limited and ultimately saturates.
As such, even if one keeps increasing $N$ and/or lowering $q$,
the relaxation rate, ${ \mR_{1} (J) }$,
saturates and stops converging further toward the discontinuous prediction
from the ${1/N}$ Landau equation (see Fig.~\ref{fig:Landau_1}).

\subsubsection{Nonlinear timescale}
\label{sec:1N_NonlinearTimescale}

Let us now be more specific on this (tentative) saturation effect.
For a given value of ${ (N,q) }$, let us therefore estimate the system's nonlinear timescale, $\Tnl$.
This time would then be the appropriate regularisation timescale, $\Treg$,
to be considered in Eq.~\eqref{eq:Landau_1N_REG}
when comparing with numerical simulations.
Heuristically, the nonlinear time, $\Tnl$,
corresponds to the duration during which the quasilinear assumption
from Eq.~\eqref{eq:linear_deltaF} holds.
In that view, let us therefore equate the contribution from the nonlinear term ${ [\delta F , \delta H] }$,
with the rate of change of the DF fluctuation, ${ \p_{t} \delta F }$, in Eq.~\eqref{eq:linear_deltaF}.
We find that ${ |\delta F| / \Tnl \!\simeq\! G \, |\delta F|^{2} }$,
where we assumed that ${ |\delta H| \!\simeq\! G \, |\delta F| }$.
Assuming that the statistics of the DF fluctuations is Poissonian~\citep[see, e.g.\@, Eq.~{(41)} in][]{Chavanis2012AA},
we can take ${ |\delta F|^{2} \!\simeq\! \gamma \Gtot }$,
with ${ \gamma \!=\! \Gtot / N }$ the vortices' individual circulations,
and ${ \Gtot }$ involving the system's active fraction.
Overall, we find the scaling
\begin{equation}
\Tnl \simeq \frac{\sqrt{N}}{q} \, \Tdyn .
\label{eq:def_Tnl}
\end{equation}
Physically, the dependence of Eq.~\eqref{eq:def_Tnl} makes sense.
Indeed, by increasing $N$, i.e.\ by reducing the level of the Poisson fluctuations,
and by lowering $q$, i.e.\ by reducing the level of collective amplification,
one increases the system's nonlinear time,
i.e.\ one allows for the quasilinear assumption to hold longer in Eq.~\eqref{eq:linear_deltaF}.

\subsubsection{Alternative view on the nonlinear timescale}
\label{sec:1N_AlternativeView}

Interestingly, taking inspiration from~\cite{Taylor+1971,Dubin2003},
one can offer another heuristic derivation of the scaling of $\Tnl$.
Rather than being driven by nonlinearities affecting
the dynamics of ${ \delta F }$, let us assume that it is the (finite-$N$ driven)
diffusion of $F$ that drives itself the broadening of the resonance condition
in Eq.~\eqref{eq:Landau_1N_REG}.
This is what we explore in more detail in Appendix~\ref{app:t}
recovering the regularised Landau equation from individual stochastic
equations of motion.
We reproduce here some of the physical insights
that can be gained from this approach.

Dropping all factors of order unity,
and keeping only the scaling dependence with respect to $N$ and $q$,
one can, heuristically, rewrite Eq.~\eqref{eq:Landau_1N_REG} as
\begin{equation}
\frac{\p F}{\p t} \propto \frac{q^{3}}{N} \, \frac{\Tnl}{1 \!+\! (\Delta \Omega \, \Tnl)^{2}} ,
\label{eq:scaling_arash}
\end{equation}
where the factor ${ q^{3} / N }$ comes from ${ \gamma F(J) F(J_{1}) }$ in Eq.~\eqref{eq:Landau_1N_REG},
${ \Delta \Omega \!=\! \Omega (J) \!-\! \Omega (J_{1}) }$
is the resonance frequency,
and we expanded the broadened Dirac, $\delta_{\Treg}$,
using Eq.~\eqref{eq:def_Lorentzian}
and imposing ${ \Tnl \!=\! \Treg }$.
Let us then assume that it is the change in $F$ described by Eq.~\eqref{eq:scaling_arash}
that is itself responsible for resonance broadening.
Hence, we may make the replacement ${ \p F / \p t \!\to\! q / \Tnl }$ in Eq.~\eqref{eq:scaling_arash},
so that it becomes
\begin{equation}
\frac{1}{\Tnl / \Tdyn} \simeq \frac{q^{2}}{N} \, \frac{\Tnl / \Tdyn}{1 \!+\! (\Delta \Omega \, \Tnl)^{2}} .
\label{eq:scaling_arash_bis}
\end{equation}
Equation~\eqref{eq:scaling_arash_bis} is a self-consistent
relation for the nonlinear time, $\Tnl$~\citep[see, e.g.\@,][]{Taylor+1971,Dubin2003}.
This self-consistent relation is explored further
in Appendix~\ref{app:Diff_extremum}
and the associated Fig.~\ref{fig:Dstar}.
Close to the extremum of the frequency profile,
we may take ${ \Delta \Omega \, \Tnl \!\to\! 0 }$,
so that Eq.~\eqref{eq:scaling_arash_bis} gives
\begin{equation}
\frac{1}{(\Tnl / \Tdyn)^{2}} \simeq \frac{q^{2}}{N} .
\label{eq:scaling_arash_ter}
\end{equation}
Ultimately, we recover the exact same scaling
as in Eq.~\eqref{eq:def_Tnl}.
Exact prefactors for this relation are obtained
in Appendix~\ref{app:Landau_from_stochastic}.
Of course, it is no surprise that both approaches lead
to the same scaling, since the kinetic equation
is, by construction, sourced by nonlinear effects,
i.e.\ one has ${ \p F / \p t \!=\! - \langle [\delta F , \delta H] \rangle }$.

\subsubsection{Numerical exploration of saturation}
\label{sec:1N_NumericalSaturation}

\begin{figure}
\begin{center}
\includegraphics[width=0.45\textwidth]{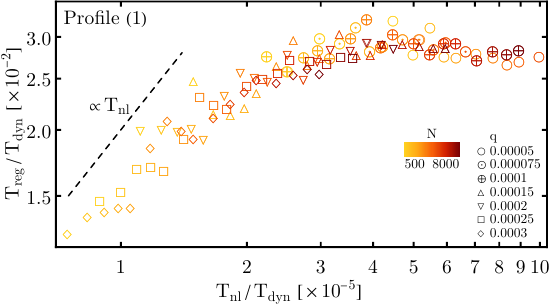}
\caption{Dependence of the regularisation time, $\Treg$ (Eq.~\ref{eq:Landau_1N_REG}),
as a function of the nonlinear time, $\Tnl$ (Eq.~\ref{eq:def_Tnl}),
using numerical simulations with different values of $N$ (different colours)
and $q$ (different symbols).
Here, $\Treg$ has been obtained following Appendix~\ref{app:1_N_prediction}
to best reproduce the relaxation rate near the frequency extremum.
Surprisingly, for large values of $\Tnl$,
we find that $\Treg$ saturates,
i.e.\ the measured relaxation rate stops converging
to the (sharp) kinetic prediction from Eq.~\eqref{eq:Landau_1N_noK}.
\label{fig:Treg_Tnl}}
\end{center}
\end{figure}

We can now use numerical simulations
to investigate the validity of the previous heuristic interpretations,
as one varies the values of ${ (N,q) }$.
For a given ${ (N,q) }$,
we can first compute the system's nonlinear time, $\Tnl$,
following Eq.~\eqref{eq:def_Tnl}.
For the same parameters, we can also rely on numerical simulations
to measure the system's relaxation rate, ${ \mR_{1} (J) }$.
Then, just like in Fig.~\ref{fig:Landau_1}, we can adjust
for the value of the regularisation time, $\Treg$ (Eq.~\ref{eq:Landau_1N_REG}),
to best reproduce the slope of the relaxation rate
near the extremum of the frequency profile (following Fig.~\ref{fig:Slope_Treg}),
i.e.\ where broadening effects play the most important role.
In Fig.~\ref{fig:Treg_Tnl}, we explore the dependence
of $\Treg$ as a function of $\Tnl$ for a large swath
of values of ${ (N,q) }$.

First, in Fig.~\ref{fig:Treg_Tnl}, 
for small values of $\Tnl$,
we find that increasing $\Tnl$, indeed, leads to an increase
in $\Treg$. Phrased differently,
in systems with strong resonance broadening,
delaying the appearance of nonlinear effects
leads to a more peaked relaxation rate near the frequency extremum.
Unfortunately, the recovery of the expected trend ${ \Treg \!\propto\! \Tnl }$
is only tentative here. This is due, in particular, to the reduced range in $\Tnl$
that was considered because of the numerical complexity
of simulating the present system of point vortices,
for a large number of pairs ${ (N,q) }$
and of independent realisations
(see Appendix~\ref{app:NumericalSimulations} for details).

Second, in Fig.~\ref{fig:Treg_Tnl}, we find
that for all systems with ${ \Tnl \!\gtrsim\! 3 \!\times\! 10^{5} \, \Tdyn }$,
the inferred regularisation timescale, $\Treg$, seems to saturate.
Phrased differently, even if one keeps increasing the value of $N$
and/or decreasing the value of $q$, the measured relaxation rate
stops converging further toward the (discontinuous) prediction
from the ${1/N}$ Landau equation (Eq.~\ref{eq:Landau_1N_noK}).
Following the numerical result from Fig.~\ref{fig:Treg_Tnl},
this suggests that the ${1/N}$ Landau equation
will not be recovered in the limit ${ N \!\to\! + \infty }$.
This is a very surprising result, and one for which
we currently have no convincing explanation.
Naturally, one could be worried that this saturation behaviour
stems from some issues in the numerical simulations.
To limit possible concerns, we performed a couple
of convergence tests, as detailed in Appendix~\ref{app:NumericalSimulations}.
To summarize, we empirically find that
\begin{equation}
\frac{\Treg}{\Tdyn} \simeq
\begin{cases}
\displaystyle 10^{-3} \frac{\sqrt{N}}{q} {} & \displaystyle \mathrm{if} \quad \frac{\sqrt{N}}{q} \lesssim 3 \!\times\! 10^{5} ,
\\[2.0ex]
\displaystyle 3 \!\times\! 10^{2} {} & \displaystyle \mathrm{if} \quad \frac{\sqrt{N}}{q} \gtrsim 3 \!\times\! 10^{5} .
\end{cases}
\label{eq:empirical_finding}
\end{equation}

Of course, the various pieces of evidence presented here
only offer some glimpse on the role of nonlinear effects
and resonance broadening near frequency extrema,
i.e.\ near the breakdown of the ${1/N}$ Landau equation.
At this stage, we can already list three possible venues
to improve upon the results presented in the present section:
(i) One should go beyond the heuristic regularisation
of Eq.~\eqref{eq:Landau_1N_REG} to derive a self-consistent
kinetic equation in that regime;
(ii) One should reproduce the effect of resonance broadening
in long-range interacting systems with much cheaper computations,
e.g.\@, with the quasiperiodic cube~\citep[see, e.g.\@,][]{Magorrian2021}
or with classical Heisenberg spins~\citep[see, e.g.\@][]{Barre+2014}.
These would be useful to check numerically the expected scaling ${ \Treg \!\propto\! \Tnl }$;
(iii) Finally, leveraging the same systems,
one should explore the relaxation of systems with much larger values of $\Tnl$
to confirm the saturation effect observed in Fig.~\ref{fig:Treg_Tnl}.
All these investigations are left for future work.

\subsection{Long-term relaxation}
\label{sec:LongTermRelaxation}

Ultimately,
the present system relaxes toward the Boltzmann distribution (Eq.~\ref{eq:def_FB}).
And, we refer to Appendix~\ref{app:Boltzmann} for details
on how this equilibrium may be determined
given some initial conditions.

\begin{figure}
\begin{center}
\includegraphics[width=0.45\textwidth]{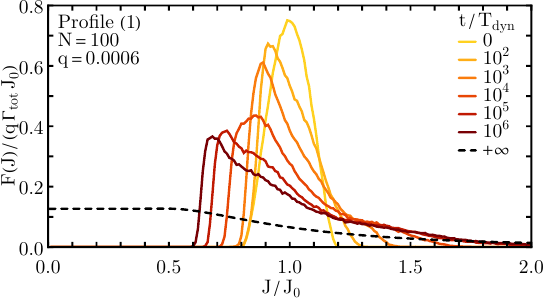}
\caption{Relaxation in $N$-body simulations of profile (1) toward the thermodynamical
equilibrium (Eq.~\ref{eq:def_FB}).
Different colours correspond to different times (here the median value over
the available realisations),
while the dashed line is the expected Boltzmann equilibrium (see Appendix~\ref{app:Boltzmann}).
Relaxation is delayed by a local kinetic blocking
induced by the absence of resonances in certain regions
of the non-monotonic frequency profile (see Fig.~\ref{fig:DF}).
\label{fig:Relaxation_1}}
\end{center}
\end{figure}

In Fig.~\ref{fig:Relaxation_1}, we use $N$-body simulations
to illustrate this long-term relaxation to the statistical equilibrium.
In that figure, one can note that the relaxation of the system
toward the central region is particularly slow.
This follows from the particular shape of the present non-monotonic frequency profile
(see Fig.~\ref{fig:DF}).
Indeed, particles within the system's
inner regions, i.e.\ with small actions, 
cannot resonantly couple to more external regions, i.e.\ to particles with larger actions.
This was responsible for the local kinetic blocking,
already visible in the ${1/N}$ Landau flux from Fig.~\ref{fig:Landau_1}.
As such, these inner regions cannot relax through $1/N$ effects
and must wait for the outer regions to diffuse,
via ${1/N^{2}}$ effects, so that non-local resonances get populated
and that the more efficient ${1/N}$ relaxation may proceed.
This lack of efficient resonant couplings
is responsible for the system's particularly slow diffusion
within the core regions.
Phrased differently, although the system's frequency profile is non-monotonic,
there still exist regions in action space that only relax,
at least initially, via ${1/N^{2}}$ effects.
Such next-order contributions are the focus of the upcoming section.

\section{${ 1/N^{2} }$ dynamics}
\label{sec:1_N2_dynamics}

\subsection{Kinetic equation}
\label{sec:1N2_kinetic_equation}

We now consider profile (2),
i.e.\ a system with a monotonic frequency profile (Fig.~\ref{fig:DF}).
In that case, the resonance condition
in the ${1/N}$ Landau equation (Eq.~\ref{eq:Landau_1N})
imposes ${ J_{1} \!=\! J }$,
hence leading to the vanishing of the crossed term ${ \bk \!\cdot\! \p F_{2} / \p \bJ }$.
This is a kinetic blocking, and there is no relaxation driven by ${1/N}$ effects.\footnote{Kinetic blocking at order ${1/N}$ also holds when accounting for collective effects,
as described by the inhomogeneous Balescu--Lenard equation.}
Relaxation is greatly hampered,
and occurs only through ${1/N^{2}}$ effects.
Placing themselves within this regime and neglecting collective effects,
\cite{Fouvry2022} derived a closed kinetic equation describing this relaxation.
The inhomogeneous ${ 1/N^{2} }$ Landau equation reads
\begin{align}
\frac{\p F (J , t)}{\p t} = \gamma^{2} \frac{\p }{\p J} \bigg[ {} & \sum_{k_{1} , k_{2}} \! (k_{1} \!+\! k_{2}) \mP \!\! \int \!\! \rd J_{1} \rd J_{2} \, \big| \Lambda_{k_{1} k_{2}} (\bJ) \big|^{2}
\nonumber
\\
\times {} & \, \deltaD \big( \bk \!\cdot\! \bO \big) \, \bk \!\cdot\! \frac{\p }{\p \bJ} F_{3} (\bJ) \bigg] ,
\label{eq:Landau_1N2}
\end{align}
with $\mP$ the Cauchy principal value acting on the integral over ${ \rd J_{1} }$.
We refer to Appendix~\ref{app:1_N2_prediction}
for the expression of the coupling coefficients, ${ |\Lambda_{k_{1} k_{2}} (\bJ)|^{2} }$,
which are independent of both $N$ and $q$.
In Eq.~\eqref{eq:Landau_1N2}, we shortened the notations with
the 3-vectors ${ \bJ \!=\! (J , J_{1} , J_{2}) }$,
${ \bO \!=\! (\Omega[J] , \Omega[J_{1}] , \Omega[J_{2}]) }$,
and ${ \bk \!=\! (k_{1} \!+\! k_{2} , - k_{1} , - k_{2}) }$,
as well as
${ F_{3} (\bJ) \!=\! F(J) F(J_{1}) F(J_{2}) }$.
At the present time, the generalisation of Eq.~\eqref{eq:Landau_1N2}
that accounts for collective effects is unknown.

Just like the ${1/N}$ Landau equation (Eq.~\ref{eq:Landau_1N}),
the ${1/N^{2}}$ Landau equation (Eq.~\ref{eq:Landau_1N2})
conserves the total circulation, momentum and energy (Eq.~\ref{eq:total_conservation}),
satisfies an $H$-theorem for the Boltzmann entropy (Eq.~\ref{eq:def_entropy}),
and vanishes exactly for the Boltzmann DF (Eq.~\ref{eq:def_FB}).

Equation~\eqref{eq:Landau_1N2} contains an intricate
resonance condition, ${ \deltaD (\bk \!\cdot\! \bO) }$, involving three particles.
Even with a monotonic frequency profile,
this resonance condition allows for non-local resonances,
i.e.\ triplets of actions, $\bJ$, that are not all identical.
As a result, no further kinetic blocking is generically to be expected.\footnote{Though, we refer to~\cite{Fouvry+2023}
for the (involved) example of a bare second-order kinetic blocking for a particular class of potentials of interaction. For the potential of interaction of point vortices
(Eqs.~\ref{eq:def_U_unsoftened} and~\ref{eq:def_U_softened}),
there is no kinetic blocking at the order $1/N^2$ and the DF relaxes
through Eq.~\eqref{eq:Landau_1N2} towards the Boltzmann DF.}
In that case, we find from Eq.~\eqref{eq:Landau_1N2} that
the relaxation time scales like
\begin{equation}
\Trelax \simeq \Tdyn \, N^{2} / q^{4} ,
\label{eq:Trelax_1N2}
\end{equation}
with respect to $N$ and $q$,
in the limit where collective effects can be neglected.
As a result, it is useful to introduce the rescaled relaxation rate
\begin{equation}
\mR_{2} (J,t) = \frac{\p N(\!<\! J,t) / \p t}{N / \Tdyn} \, \frac{N^{2}}{q^{4}} ,
\label{eq:def_R2}
\end{equation}
using the same notation as in Eq.~\eqref{eq:def_R1}.
In that case, the ${1/N^{2}}$ Landau equation (Eq.~\ref{eq:Landau_1N2})
predicts that ${ \mR_{2} (J) }$ should be independent of both $N$ and $q$.

\subsection{Linear stability}
\label{sec:1N2_Stability}

\begin{figure}
\begin{center}
\includegraphics[width=0.45\textwidth]{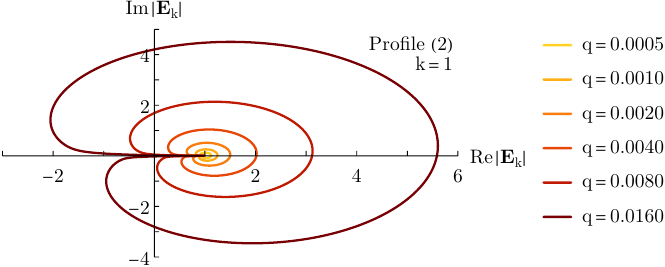}
\caption{Same as Fig.~\ref{fig:Nyquist_1} but for profile (2).
Here, for all values of $q$, the system is always found to be linearly stable.
\label{fig:Nyquist_2}}
\end{center}
\end{figure}

The scaling from Eq.~\eqref{eq:Trelax_1N2} only holds
in the limit of weak collective amplification.
Once again, we can use linear response theory
to determine an appropriate range of active fraction
for which collective effects are negligible.
This is illustrated in Fig.~\ref{fig:Nyquist_2} for profile (2)
using Nyquist contours.
Interestingly, in this figure, we find that the geometry of the Nyquist contours
is such that, whatever the considered value of $q$, the contour never encloses
the origin of the complex plane. Phrased differently,
contrary to profile (1),
here for profile (2), whatever the system's active fraction,
the system's initial distribution is linearly stable.
Of course, linear stability is not synonymous with weak collective amplification.
Indeed, in Fig.~\ref{fig:Nyquist_2}, we find that for ${ q \!\gtrsim\! 0.004 }$,
the Nyquist contour gets extremely close to the origin,
i.e.\ there exist real frequencies for which the system amplifies very significantly
perturbations. In practice, in all the upcoming simulations,
we limit ourselves to simulations with
${ q \!\leq\! 0.0007 }$,
so that Eq.~\eqref{eq:Landau_1N2} applies.

\subsection{Small-scale contributions}
\label{sec:1N2_SmallScale}

As for the ${1/N}$ case, let us first focus
on the contributions from small scales.
Indeed, Eq.~\eqref{eq:Landau_1N2} involves an infinite sum over the resonance numbers
${ (k_{1} , k_{2}) }$, i.e.\ one ought to ensure
that the limits ${ k_{1} , k_{2} \!\to\! + \infty }$ are meaningful.

\begin{figure}
\begin{center}
\includegraphics[width=0.45\textwidth]{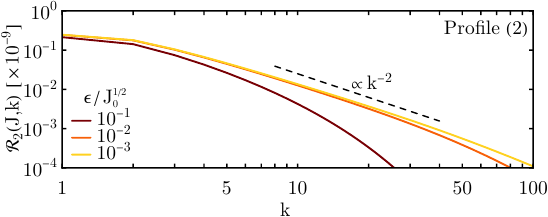}
\caption{Same as Fig.~\ref{fig:Coulomb_1},
but for the ${1/N^{2}}$ Landau equation (Eq.~\ref{eq:Landau_1N2})
and profile $(2)$ (Eq.~\ref{eq:def_R2k}).
For small $k$, i.e.\ scales larger than $\eps$,
we recover the expected scaling in $k^{-2}$.
\label{fig:Coulomb_2}}
\end{center}
\end{figure}

Following Eq.~\eqref{eq:Landau_1N2},
we decompose the relaxation rate from Eq.~\eqref{eq:def_R2} as
\begin{equation}
\mR_{2} (J) = \sum_{k > 0} \mR_{2} (J , k) ,
\label{eq:def_R2k}
\end{equation}
where we introduced
\begin{equation}
\mR_{2} (J , k) = \sum_{\mathclap{\substack{k_{1} , k_{2} \\ \mathrm{Max} [|k_{1}|,|k_{2}|] = k}}} \mR_{2} (J , k_{1} , k_{2}).
\label{eq:def_R2kk}
\end{equation}
In Fig.~\ref{fig:Coulomb_2}, we illustrate the dependence of ${ k \!\mapsto\! \mR_{2} (J \!=\! 1.1 \, J_{0} , k) }$.

From that figure, one can make a couple of observations:
(i) For $k$ small enough, one finds ${ \mR_{2} (J , k) \!\propto\! k^{-2} }$.
This is in agreement with the decay rate that can directly
be inferred from Eq.~\eqref{eq:Landau_1N2},
as detailed in Appendix~\ref{app:1_N2_prediction}.
(ii) As one increases the softening length, $\eps$,
the contribution to the relaxation rate decays faster with $k$.
As expected, introducing softening leads to the vanishing
of the contribution from small scales.
Overall, we find that Eq.~\eqref{eq:Landau_1N2}
does not suffer from any small scale divergence (even in the absence of softening),
when applied to the dynamics of point vortices.

\subsection{Relaxation rate}
\label{sec:1N2_RelaxationRate}

\begin{figure}
\begin{center}
\includegraphics[width=0.45\textwidth]{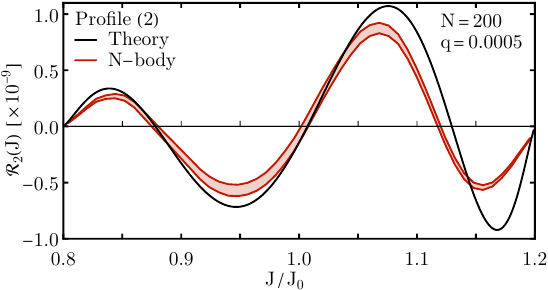}
\caption{Same as Fig.~\ref{fig:Landau_1}
but for profile (2).
For that profile, relaxation is driven by ${1/N^{2}}$ effects (Eq.~\ref{eq:Landau_1N2}),
and hence is much delayed.
The origin of the mismatch for large actions remains unclear.
\label{fig:Landau_2}}
\end{center}
\end{figure}

We are now set to compare the kinetic prediction from Eq.~\eqref{eq:Landau_1N2}
with measurements in $N$-body simulations.
This is presented in Fig.~\ref{fig:Landau_2}.
In that figure, we find a welcome agreement
between the ${1/N^{2}}$ prediction and the numerical measurements.
This is one of the main result of the present work,
i.e.\ a quantitative validation of the ${1/N^{2}}$ kinetic theory of point vortices.
One should point out the presence of a mismatch for ${ J / J_{0} \!\gtrsim\! 1.1 }$.
Such a disagreement could stem
from contributions associated with collective effects,
since Eq.~\eqref{eq:Landau_1N2} neglects them altogether.
We point out that all previous explicit applications of the ${1/N^{2}}$ kinetic equation
have always uncovered a mismatch,
see, in particular, fig.~{2} in~\cite{Fouvry+2019} for the HMF model,
fig.~{1} in~\cite{Fouvry+2020} for the ring model, and
fig.~{1} in~\cite{Fouvry2022} for Heisenberg spins.
A next step to clarify this effect
would be to generalise Eq.~\eqref{eq:Landau_1N2}
to the case with collective effects,
i.e.\ to derive, for the first time, the ${1/N^{2}}$ Balescu--Lenard equation.

\begin{figure}
\begin{center}
\includegraphics[width=0.49\textwidth]{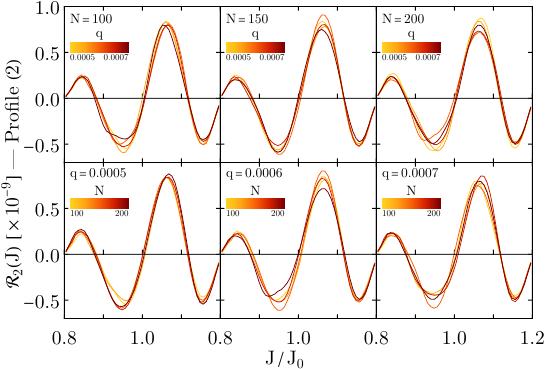}
\caption{Same as Fig.~\ref{fig:Flux_Grid_2},
but for profile (2).
As one varies $N$ and/or $q$,
the (rescaled) relaxation rate, ${ \mR_{2} (J) }$ (Eq.~\ref{eq:def_R2}),
remains, in essence, unchanged.
\label{fig:Flux_Grid_2}}
\end{center}
\end{figure}

Building upon the previous case of profile (1),
one could be worried that resonance broadening effects
could be polluting the measurement from Fig.~\ref{fig:Landau_2}.
In Fig.~\ref{fig:Flux_Grid_2},
we repeat the measurement of the relaxation rate, ${ \mR_{2} (J) }$,
for various values of $N$ and $q$.
In that figure, contrary to Fig.~\ref{fig:Flux_Grid_1},
we observe no clear trend with respect to modifications
of $N$ and/or $q$.
As a result, for the present monotonic frequency profile,
there does not seem to be any clear signature
from nonlinear effects
that would cause a significant broadening
of the three-body resonance condition from Eq.~\eqref{eq:Landau_1N2}.

\section{Conclusions}
\label{sec:Conclusion}

In this paper, we investigated the long-term relaxation
of long-range interacting point vortices in ${2D}$ hydrodynamics
within axisymmetric configurations.
We showed how the generic kinetic theory
of inhomogeneous long-range interacting systems
can readily be applied to point vortices.
In that context, we emphasised the existence of two regimes of relaxation:
(i) if the system's mean frequency profile is non-monotonic,
the distribution of point vortices relaxes through ${1/N}$ effects.
In the limit of negligible collective amplification,
this is governed by the ${1/N}$ inhomogeneous Landau equation
(Eq.~\ref{eq:Landau_1N});
(ii) if the mean frequency profile is monotonic,
the system is then submitted to a kinetic blocking,
which prevents ${1/N}$ effects from driving any relaxation.
In that case, neglecting once again collective effects,
the system's relaxation is described by the ${1/N^2}$ inhomogeneous Landau equation
(Eq.~\ref{eq:Landau_1N2}).

For both regimes of relaxation, we reported on the agreement
between the prediction from kinetic theories and measurements in direct numerical simulations.
Using these same simulations, we also recovered the expected scalings
with respect to the total number of particles and the system's active fraction.
Finally, we emphasised the signature of nonlinear resonance broadening
near the extremum of the frequency profile.

Naturally, the present work is only one small step toward
building a finer understanding of long-term relaxation.
Let us conclude this work by mentioning a few possible venues
for future investigations.

\textit{${1/N^{2}}$ Balescu--Lenard equation}.
At order ${1/N}$, when collective effects cannot be neglected,
the Landau equation becomes the Balescu--Lenard equation~\citep{Heyvaerts2010,Chavanis2012AA}.
In the presence of a kinetic blocking,
the ${1/N^{2}}$ version of the Balescu--Lenard equation is currently unknown.
Although this is no easy challenge,
it would be worthwhile to derive such a generalisation,
to investigate the impact of collective effects in kinetic blockings.

\textit{Resonance broadening}. Of course, it would be interesting
to investigate further the effect of resonance broadening unveiled
in Fig.~\ref{fig:Landau_1}. On the numerical front, one could revisit this problem
using $N$-body simulations in a simpler toy model.
For example, the homogeneous Hamiltonian Mean Field model~\cite{Antoni+1995}
with the kinetic energy, ${ \tfrac{1}{2} v^{2} }$,
changed for ${ \tfrac{1}{3} v^{3} }$, hence ensuring a non-monotonic mean frequency profile,
similar to the one from profile (1) here (see Fig.~\ref{fig:DF}).
Similarly, the case of classical Heisenberg spins could also prove insightful~\citep[see, e.g.\@][]{Barre+2014}.
On the theoretical front, going beyond the heuristic regularisation
from Eq.~\eqref{eq:Landau_1N_REG}
is particularly challenging, in particular to ensure the expected conservation laws
and $H$-theorem. One possible venue could be to explore
systematic methods stemming from renormalisation theory~\cite[see, e.g.\@,][ and references therein]{Krommes2002}.

\textit{Large deviations}. We focused our interest
on the ensemble-averaged evolution.
Going beyond the mean dynamics,
one should also investigate the dispersion of independent realisations
around that mean evolution.
This is the promise of large deviation theory applied to kinetic equations~\cite{Feliachi+2022,Feliachi+2024},
though its quantitative comparison with $N$-body simulations
remains to be done.

\textit{Marginal stability}. In the presence of a monotonic frequency profile,
an axisymmetric distribution of point vortices is always submitted
to a kinetic blocking, independently of the strength of collective effects.
Yet, as a system nears marginal stability,
i.e.\ as the oscillation of the damped modes take longer and longer to fade away,
mode-particle couplings are expected to get more and more efficient
at driving the system's relaxation~\citep[see, e.g.\@,][]{Rogister+1968,Hamilton+2020}.
It would be interesting to investigate the impact
that such ``quasilinear'' contributions can have on kinetic blockings.

\textit{Spectrum of circulations}. Here, we considered systems
with a single population of vortices,
sharing all the exact same individual circulation.
In the view of better understanding segregation processes,
one should investigate systems with different circulations,
in particular different signs~\citep[see, e.g.\@,][]{Dubin2003,Chavanis2023}.

\begin{acknowledgments}
This work is partially supported by the grant Segal ANR-19-CE31-0017 
of the French Agence Nationale de la Recherche
and by the Idex Sorbonne Universit\'e.
This research was supported in part by grant NSF PHY-2309135
to the Kavli Institute for Theoretical Physics (KITP).
We warmly thank S.\ Flores and A.\ El Rhirhayi for detailed remarks and suggestions.
We are grateful to S.\ Rouberol for the smooth running
of the Infinity cluster where the simulations were performed.
\end{acknowledgments}

\vskip 1cm

\appendix

\section{Background profiles}
\label{app:BackgroundProfiles}

In this Appendix, we detail our prescription for the external potentials
introduced in Eq.~\eqref{eq:def_Uext}.

The mean self-consistent potential generated by the point vortices reads
\begin{align}
H_{\eps} [F] {} & = \!\! \int \!\! \rd \brp \, U_{\eps} (\br , \brp) \, F (\rp)
\nonumber
\\
{} & = 2 \pi \!\! \int \!\! \rd \Jp \, U_{\eps, k = 0} (J , \Jp) \, F (\Jp) ,
\label{eq:calc_Heps}
\end{align}
with ${ F (\Jp) }$ the \DF\ from Eq.~\eqref{eq:DF_LOC},
and the Fourier-transformed (softened) interaction potential, ${ U_{k,\eps} (J , \Jp) }$
following Eq.~\eqref{eq:def_U_soft}.
Fortunately, this integral, as well as the associated orbital frequency,
can be explicitly performed,
using symbolic computing (\texttt{Mathematica}),
even in the presence of softening.
We implemented these expressions in the $N$-body simulations (see Appendix~\ref{app:NumericalSimulations}).

For profile (1), we assume that it is generated
by the background \DF\@
\begin{equation}
\Fb (J) = \Theta [J \!\geq\! \Jb] \, \frac{\Gtotb}{\pi \sigmab} \, \frac{1}{[1 + |J \!-\! \Jb| / \sigmab]^{3}} ,
\label{eq:DF_Profile_1}
\end{equation}
with ${ \Gtotb \!=\! \Gtot }$ the background total circulation.
Here, $\Jb$ is the shift of the background \DF\@, while $\sigmab$ is its width.
Using once again Eq.~\eqref{eq:calc_Heps},
one can compute explicitly the associated, unsoftened, potential.
Although somewhat lengthy, this expression can be obtained
using symbolic computing.
In practice, in the $N$-body simulations,
we imposed ${ \Jb / J_{0} \!=\! \half }$ and ${ \sigmab / J_{0} \!=\! 1 }$,
for which the generated frequency profile is non-monotonic
in the vicinity of the active \DF\@, see Fig.~\ref{fig:DF}.

For profile (2), we consider Eq.~\eqref{eq:DF_Profile_1}
in the limit ${ \Jb \!=\! 0 }$. In that case, the (unsoftened) mean Hamiltonian reads
\begin{equation}
H_{0} (J) = \frac{G \Gtotb}{4 \pi} \, \frac{1 - (1 \!+\! J / \sigmab) \, \ln [2 (J \!+\! \sigmab)]}{1 \!+\! J / \sigmab} ,
\label{eq:H_Profile_2}
\end{equation}
while the mean frequency is ${ \Omega_{0} (J) \!=\! \Omegab \, g (J / \sigmab) }$
with the frequency scale ${ \Omegab \!=\! - G \Gtotb / (2 \pi \sigmab) }$
and the dimensionless monotonic function
\begin{equation}
g(x) = \frac{1 \!+\! \half x}{(1 \!+\! x)^{2}} .
\label{eq:g_Profile_2}
\end{equation}
Such a model is a pleasant choice as it allows
for the resonance condition in Eq.~\eqref{eq:Landau_1N2}
to be explicitly solved. Indeed, for ${ 0 \!<\! \varpi \!\leq\! 1 }$,
one has
\begin{equation}
g (x) = \varpi
\;\; \Longrightarrow \;\;
x = \frac{1 \!-\! 4 \varpi \!+\! \sqrt{1 \!+\! 8 \varpi}}{4 \varpi} .
\label{eq:solve_g}
\end{equation}
This greatly eases the numerical implementation
of the ${ 1/N^{2} }$ kinetic equation.
In practice, in the $N$-body simulations,
we imposed ${ \sigmab / J_{0} \!=\! 1 }$.

\section{Interaction potential}
\label{app:Interaction potential}

In the absence of any softening,
the pairwise interaction potential (Eq.~\ref{eq:def_U_unsoftened})
reads
\begin{equation}
U (\br , \brp) = - \frac{G}{2 \pi} \, \ln |\br \!-\! \brp| . 
\label{eq:def_U_unsoftened_repeat}
\end{equation}
Its Fourier expansion in angle is
\begin{equation}
U (\br , \brp) = \sum_{k = - \infty}^{+ \infty} U_{k} (J, \Jp) \, \re^{\ri k (\theta - \thetap)} ,
\label{eq:def_Fourier_U}
\end{equation}
with ${ U_{k} (J , \Jp) \!=\! U_{k} (r[J] , \rp[\Jp]) }$ following~\citep[see eq.~{(A15)} in][]{Chavanis2012Vortex}
\begin{subequations}
\begin{align}
U_{0} (r , \rp) {} & = - \frac{G}{2 \pi} \, \ln (\rmax) ,
\label{eq:U0_generic}
\\
U_{k} (r , \rp) & \, = \frac{G}{4 \pi} \frac{1}{|k|} \, \bigg( \frac{\rmin}{\rmax} \bigg)^{|k|} .
\label{eq:Uk_generic}
\end{align}
\label{eq:Uk_generic_full}\end{subequations}
In that expression, we introduced ${ \rmin \!=\! \Min[r , \rp] }$ and ${ \rmax \!=\! \Max [r , \rp] }$.
Using the relation ${ J \!=\! \half r^{2} }$,
it is then straightforward to obtain the coupling coefficient in action space,
${ U_{k} (J , \Jp) }$.

To soften the interaction on small scales,
we transform Eq.~\eqref{eq:def_U_unsoftened_repeat} into
\begin{equation}
U (\br , \brp) = - \frac{G}{2 \pi} \, \ln \big( \big[ |\br \!-\! \brp|^{2} \!+\! \eps^{2} \big]^{1/2} \big) ,
\label{eq:def_U_soft}
\end{equation}
with $\eps$ some fixed softening length.
Following the same approach as in appendix~{B} of~\cite{Weinberg1986},
the Fourier expansion from Eq.~\eqref{eq:Uk_generic_full} becomes
\begin{subequations}
\begin{align}
U_{0} (r , \rp) {} & = - \frac{G}{2 \pi} \, \ln (r_{\alpha}) ,
\label{eq:U0_soft}
\\
U_{k} (r, \rp) {} & = \frac{G}{4 \pi} \frac{1}{|k|} \bigg( \frac{r_{\beta}}{r_{\alpha}} \bigg)^{|k|} ,
\label{eq:Uk_soft}
\end{align}
\label{eq:Uk_generic_soft_full}\end{subequations}
where we introduced the two radii
\begin{subequations}
\begin{align}
r_{\alpha} {} & = \bigg[ \half \bigg( \big\{ \big( (r \!+\! \rp)^{2} \!+\! \eps^{2} \big) \, \big( (r \!-\! \rp)^{2} \!+\! \eps^{2} \big) \big\}^{1/2}
\label{eq:def_ra}
\\
{} & \quad\quad\quad + r^{2} \!+\! r^{\prime 2} \!+\! \eps^{2} \bigg) \bigg]^{1/2} ,
\nonumber
\\
r_{\beta} {} & = \frac{r \rp}{r_{\alpha}} .
\label{eq:def_rb}
\end{align}
\label{eq:def_rarb}\end{subequations}

\section{Linear response theory}
\label{app:LinearResponse}

\textit{Balescu--Lenard equation}. When accounting for collective effects,
the inhomogeneous Landau equation (Eq.~\ref{eq:Landau_1N})
becomes the inhomogeneous \BL\ equation~\citep{Heyvaerts2010,Chavanis2012AA}.
It has the exact same functional form
except that the bare coupling coefficients, ${ U_{k} (J , \Jp) }$,
become the (frequency-dependent) dressed ones, ${ \Ud_{k} (J , \Jp ; \omega) }$,
to be used in Eq.~\eqref{eq:def_Lambda1}.

\textit{Self-consistent relation}. The dressed coupling coefficients satisfy the self-consistent relation~\citep[see, e.g.\@,][]{Chavanis2012AA}
\begin{align}
\Ud_{k} (J , \Jp ; {} & \omega) = U_{k} (J , \Jp)
\label{eq:self_Ud}
\\
+ {} & \, 2 \pi \!\! \int \!\! \rd \Jpp \frac{k \p F / \p \Jpp}{(k \Omegapp - \omega)_{-}} \, U_{k} (J , \Jpp) \, \Ud_{k} (\Jpp , J ; \omega) ,
\nonumber
\end{align}
with ${ \Omegapp \!=\! \Omega (\Jpp) }$.
We also introduced the regularised complex Hilbert kernel~\citep[see, e.g.\@,][and references therein]{Fouvry+2022}
\begin{equation}
\frac{1}{z_{-}} = \lim\limits_{T \to + \infty} \frac{1 - \re^{- \ri z T}}{z} .
\label{eq:def_Hilbert}
\end{equation}
The dressed coupling coefficients, ${ \Ud_{k} (\omega) }$, capture the strength
of the collective amplification in the system.

\textit{Matrix rewriting}. In inhomogeneous systems, ${ \Ud_{k} (\omega) }$, is typically evaluated
using Kalnajs' basis method~\citep{Kalnajs1976}.
Here, we resort to a simpler method
that frees us from having to construct tailored basis elements,
as we now detail.

First, in Eq.~\eqref{eq:self_Ud},
we note that
both the resonance number, $k$, and the frequency, $\omega$,
are external labels of the integral over ${ \rd \Jpp }$.
Let us therefore assume that we can decompose the action domain
into a finite number of bins of length ${ \Delta J }$.
This provides us then with a list of action, ${ \{ J_{i} \}_{i} }$,
around which each bin is centred. We now introduce the matrices,
$\bU_{k}$ and ${ \bUd_{k} (\omega) }$, via
\begin{subequations}
\begin{align}
\big( \bU_{k} \big)_{ij} {} & = U_{k} (J_{i} , J_{j}) ,
\label{eq:def_bU}
\\
\big( \bUd_{k} (\omega) \big)_{ij} {} & = \Ud_{k} (J_{i} , J_{j} ; \omega) .
\end{align}
\label{eq:def_mat}\end{subequations}
Approximating the integral from Eq.~\eqref{eq:self_Ud}
with the midpoint rule, it becomes the matrix relation
\begin{equation}
\bUd_{k} (\omega) = \bU_{k} + \bU_{k} \, \bM_{k} (\omega) \, \bUd_{k} (\omega) ,
\label{eq:self_mat}
\end{equation}
where we introduced the diagonal matrix
\begin{equation}
\big( \bM_{k} (\omega) \big)_{ij} = \delta_{ij} \, 2 \pi \, \Delta J \, \frac{k \p F / \p J_{i}}{(k \Omega (J_{i}) - \omega)_{-}} .
\label{eq:def_bM}
\end{equation}
After manipulation, Eq.~\eqref{eq:self_mat} finally becomes
\begin{equation}
\bE_{k} (\omega) \, \bUd_{k} (\omega) = \bU (k) ,
\label{eq:ampl_mat}
\end{equation}
or equivalently
\begin{equation}
\bUd_{k} (\omega) = \bE_{k} (\omega)^{-1} \, \bU (k) ,
\end{equation}
where we introduced the ``dielectric'' matrix
\begin{equation}
\bE_{k} (\omega) = \bI - \bU (k) \, \bM_{k} (\omega) ,
\label{eq:def_bE}
\end{equation}
with $\bI$ the identity matrix.
Heuristically, Eq.~\eqref{eq:ampl_mat} should be interpreted as
${ \Ud \!\sim\! U / E }$, hence describing collective dressing.
In particular, for a given resonance number, $k$,
a system supports a mode at frequency $\omegaM$ if one has
\begin{equation}
\det \big[ \bE_{k} (\omegaM) \big] = 0 .
\label{eq:def_mode}
\end{equation}
Importantly, the present approach
(i) does not require any potential/density basis elements;
(ii) can readily be applied to softened pairwise interactions.
Since the approach from this Appendix directly computes the dressed coupling
coefficients, $\Ud$, it can readily be used to predict the \BL\ relaxation rate~\citep{Heyvaerts2010,Chavanis2012AA}.
This provides an alternative to the traditional basis method~\citep{Kalnajs1976}
used, for example, in the \BL\ predictions of~\cite{Roule+2022}.

\textit{Nyquist contours}. In Figs.~\ref{fig:Nyquist_1} and~\ref{fig:Nyquist_2},
the Nyquist contours represent, in the complex plane,
the values of the function
${ \omegaR \!\in\! \mathbb{R} \!\to\! \det [\bE_{k} (\omegaR \!+\! \ri \eta)] \!=\! |\bE_{k} (\omegaR \!+\! \ri \eta)| }$.
Phrased differently, these figures show the imaginary part
of ${ |\bE_{k} (\omegaR \!+\! \ri \eta )| }$ as a function of its real part,
as one varies $\omegaR$.
In practice, (i) we used $500$ points to sample the action domain in Eq.~\eqref{eq:def_mat},
and (ii) we regularised Eq.~\eqref{eq:def_bM} with the small positive imaginary part
${ \eta \!=\! 10^{-3} \, \Omega (J_{0}) }$, so that
${ 1/(\omegaR \!+\! \ri \eta)_{-} \!\to\! 1/(\omegaR \!+\! \ri \eta) }$.
We checked that varying the number of sampling points
or the value of $\eta$
had no impact on the plotted Nyquist curves.

\section{${ 1/N }$ kinetic theory}
\label{app:1_N_prediction}

\textit{${1/N}$ Landau equation}. Following~\cite{Chavanis2012AA}, in the absence of collective effects,
the inhomogeneous Landau equation (Eq.~\ref{eq:Landau_1N})
satisfies
\begin{equation}
\big| \Lambda_{k} (\bJ) \big|^{2} = 2 \pi^{2} \, \big| U_{k} (J , J_{1}) \big|^{2} ,
\label{eq:def_Lambda1}
\end{equation}
where ${ U_{k} (J , J_{1}) }$ is the Fourier transformed interaction potential
(see Appendix~\ref{app:Interaction potential}).
Performing the sum over $k$, one readily obtains Eq.~\eqref{eq:Landau_1N_noK}.
For the unsoftened interaction potential,
the sum can be carried out explicitly~\citep[see, e.g.\@, appendix B in][]{Chavanis2010}.

\textit{Small-scale contributions}. In Fig.~\ref{fig:Coulomb_1},
we present the respective contribution of small scale resonances,
i.e.\ large $k$,
to the ${1/N}$ Landau relaxation rate.
Let us recover the main features of that figure
directly from Eq.~\eqref{eq:Landau_1N}.
For ${ \Jp \!\to\! J }$, Eq.~\eqref{eq:Uk_generic} gives
${ U_{k} (J , \Jp) \!\simeq\! k^{-1} }$,
so that Eq.~\eqref{eq:def_Lambda1} gives ${ |\Lambda_{k} (\bJ)|^{2} \simeq k^{-2} }$.
When used in Eq.~\eqref{eq:def_Lambdatot},
we find from Eq.~\eqref{eq:Landau_1N_noK}
that each scale contributes like $k^{-1}$.
In practice, this scaling of the relaxation rate only holds for $k$ small enough,
before the difference between the two resonating actions $J$ and $J_{1}$
leads a decay faster than $k^{-1}$ in Eq.~\eqref{eq:Uk_generic}.
For the diffusion coefficient from Eq.~\eqref{eq:def_D2},
because local resonances always contribute to it,
i.e.\ ${ J_{1} \!=\! J }$ in Eq.~\eqref{eq:def_D2},
the exact match of the two resonating actions
allows for the scaling in $k^{-1}$ to hold throughout,
as visible in Eq.~\eqref{eq:Uk_generic}.
This is responsible for a small-scale logarithmic divergence
of the diffusion coefficient,
in systems driven by the unsoftened interaction
from Eq.~\eqref{eq:def_U_unsoftened} (see Appendix~\ref{app:reglog}).
As could have been expected, increasing the considered softening length
accelerates the decay of the small-scale contributions.
These various trends are further discussed in Section~\ref{sec:1N_SmallScale}
and recovered in Fig.~\ref{fig:Coulomb_1}.

\textit{Computing the kinetic prediction}. In practice, to evaluate Eq.~\eqref{eq:Landau_1N_noK},
one invokes the relation
\begin{equation}
\deltaD \big[ \Omega (J) \!-\! \Omega (J_{1}) \big] = \sum_{\Jr} \frac{\deltaD (J \!-\! \Jr)}{|\rd \Omega / \rd \Jr|} ,
\label{eq:solution_Dirac}
\end{equation}
where the $\Jr$ satisfy ${ \Omega (\Jr) \!=\! \Omega (J) }$.
Used in Eq.~\eqref{eq:Landau_1N_noK},
we get
\begin{align}
\frac{\p F (J , t)}{\p t} = \gamma \frac{\p }{\p J} \bigg[ \sum_{\Jr} {} & \frac{\big| \Lambdatot (J , \Jr) \big|^{2}}{|\rd \Omega / \rd \Jr |} 
\label{eq:Landau_1N_shortest}
\\
\times {} & \bigg( F (\Jr) \frac{\p F}{\p J} \!-\! F (J) \frac{\p F}{\p \Jr} \bigg) \bigg] .
\nonumber
\end{align}
We implemented this expression
to obtain Fig.~\ref{fig:Landau_1},
limiting the sum over $k$ in Eq.~\eqref{eq:def_Lambdatot}
to ${ |k| \!\leq\! 100 }$.
We used the same approach to compute
the ${1/N}$ Landau diffusion coefficient from Eq.~\eqref{eq:def_D2}.

\textit{Computing the regularised prediction}.
To compute prediction from the ``regularised'' ${ 1/N }$ Landau equation
(Eq.~\ref{eq:Landau_1N_REG}),
we evaluated the integral over ${ \rd J_{1} }$ therein
using a simple midpoint rule with 1\,000 nodes
within the interval ${ J_{0} \!-\! \sigma_{0} \!\leq\! J_{1} \!\leq\! J_{0} \!+\! \sigma_{0} }$,
limiting, once again, the sum over $k$ in Eq.~\eqref{eq:def_Lambdatot}
to ${ |k| \!\leq\! 100 }$.
We proceeded similarly to evaluate the regularised
diffusion coefficient from Eq.~\eqref{eq:D2_1N_REG}.

\begin{figure}
\begin{center}
\includegraphics[width=0.45\textwidth]{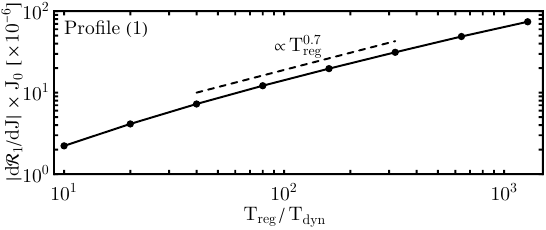}
\caption{Dependence of the predicted central slope of the relaxation rate
with respect to the action (see, e.g.\@, Fig.~\ref{fig:Flux_Treg_1}),
as a function of the regularisation time, $\Treg$, used in Eq.~\eqref{eq:Landau_1N_REG}.
\label{fig:Slope_Treg}}
\end{center}
\end{figure}

\textit{Inferring the regularisation time}. In Fig.~\ref{fig:Treg_Tnl}, we compare
the value of the nonlinear time, $\Tnl$ (Eq.~\ref{eq:def_Tnl}),
with the regularisation time, $\Treg$,
inferred from $N$-body simulations.
To obtain this figure, we proceeded as follows.
First, we computed the regularised relaxation rate
from Eq.~\eqref{eq:Landau_1N_REG}
for ${ \Treg / \Tdyn \!=\! 10 \!\times\! 2^{n} }$
with ${ 0 \!\leq\! n \!\leq\! 7 }$.
Then, for each such prediction,
we performed a linear fit of the relaxation rate within the domain
${ 1.022 \!\leq\! J / J_{0} \!\leq\! 1.042 }$,
the action domain centred around the change of sign
of the relaxation rate.
These measurements are presented in Fig.~\ref{fig:Slope_Treg}.
In that figure, we find a simple power-law dependence
linking the regularisation time, $\Treg$,
with the central slope of the relaxation rate as a function of the action.
Finally, we used the measurements from Eq.~\eqref{fig:Slope_Treg}
to infer the ``best'' regularisation time presented in Fig.~\ref{fig:Treg_Tnl}.

\section{Boltzmann equilibrium}
\label{app:Boltzmann}

The present systems have three global invariants.
Following Eq.~\eqref{eq:total_conservation}, these are
(i) the total circulation $\Gamma$;
(ii) the total momentum $L$;
and (iii) the total energy $E$.
In practice, since we consider systems with small active fraction, $q$,
for simplicity, we may approximate the total energy
by limiting ourselves only to the axisymmetric external potential, $H_{0}$,
introduced in Eq.~\eqref{eq:def_Uext}.
Doing so, we neglect the ``self-gravitating'' contribution
from the particles' pairwise couplings,
and the total energy (Eq.~\ref{eq:total_energy}) simply becomes
${ E \!=\! \!\int\! \rd J \, H_{0} (J) \, F(J,t) }$.

As a result of these conservations,
the associated Boltzmann equilibria generically read (Eq.~\ref{eq:def_FB})
\begin{equation}
\FB (J) = \alpha \, \re^{- \beta H_{0} (J) + \gamma J} ,
\label{eq:def_FB_app}
\end{equation}
where ${ (\alpha , \beta , \gamma) \!=\! \bX }$ are three parameters to be found
to comply with the conservation of ${ (\Gamma , L , E) \!=\! \bY }$.
Phrased differently, given some initial \DF\@, ${ F_{0} \!=\! F (t \!=\! 0) }$,
the associated long-term thermodynamical Boltzmann equilibrium
is specified by the set of parameters, $\bX$,
such that ${ \bY [\FB(\bX)] \!=\! \bY [F_{0}] }$.

Such a self-consistent relation can be solved using Newton's algorithm.
Given some guess $\bX^{(n)}$, we perform the iteration
\begin{equation}
\bX^{(n+1)} = \bX^{(n)} - \msJ^{-1} [\bX^{(n)}] \, \big( \bY [\bX^{(n)}] \!-\! \bY_{0} \big) ,
\label{eq:Newton_iteration}
\end{equation}
with ${ \bY_{0} \!=\! \bY [F_{0}] }$.
In that expression, we also introduced the Jacobian matrix
\begin{equation}
\big( \msJ [\bX] \big)_{ij} = \frac{\p Y_{i} [\bX]}{\p X_{j}} .
\label{eq:def_Jac}
\end{equation}
One can then iterate Eq.~\eqref{eq:Newton_iteration}
up to convergence, i.e.\ up to ${ \max_{i} |X_{i}^{(n+1)} \!-\! X_{i}^{(n)} | \!\leq\! \eps }$,
with, here, ${ \eps \!=\! 10^{-12} }$.

To obtain the Boltzmann \DF\ represented in Fig.~\ref{fig:Relaxation_1},
we computed numerically
the integrals defining the invariants $\bY$ (Eq.~\ref{eq:total_conservation})
as well as the Jacobian $\msJ$ (Eq.~\ref{eq:def_Jac})
using the midpoint rule
with $10^{5}$ nodes within the range
${ 0 \!\leq\! J / J_{0} \!\leq\! 10^{3} }$.
The initial guess was fixed to ${ \bX^{(0)} \!=\! [q \Gtotb / (2 \pi) , 0 , 0] }$.
Convergence was reached in 20 iterations,
leading to the final value
${ \bX^{(+\infty)} \!\simeq\! [ 8.88 \!\times\! 10^{-4} / \Gtotb , -40.0 / (G \Gtotb) , 6.26 \!\times\! 10^{-3} / J_{0}] }$.

\section{${ 1/N^{2} }$ kinetic theory}
\label{app:1_N2_prediction}

\textit{${1/N^{2}}$ Landau equation}. Following~\cite{Fouvry2022},
the coupling coefficients of the inhomogeneous ${1/N^{2}}$ Landau equation
read
\begin{align}
\big| \Lambda_{k_{1} k_{2}} (\bJ) \big|^{2} \!=\! 2 \pi^{3} \bigg| \frac{\big[ \Omega (J) \!-\! \Omega (J_{1}) \big] \mU^{(1)}_{k_{1} k_{2}} \!(\bJ) \!+\! k_{2} \mU^{(2)}_{k_{1} k_{2}} \!(\bJ)}{k_{1} (k_{1} \!+\! k_{2}) \big[ \Omega (J) \!-\! \Omega (J_{1}) \big]^{2}} \bigg|^{2} ,
\label{eq:exp_Lambda2}
\end{align}
with the coupling functions, $\mU^{(1)}$ and $\mU^{(2)}$,
defined in Appendix~{B} of~\cite{Fouvry2022}.
In Eq.~\eqref{eq:Landau_1N2}, the sum over the resonance vectors ${ (k_{1} , k_{2}) }$
is restricted to those such that $k_{1}$, $k_{2}$, as well as ${ k_{1} \!+\! k_{2} }$
are all non-zero.
Importantly, we emphasise that ${ | \Lambda_{k_{1} k_{2}} (\bJ) |^{2} }$
is independent of $N$ and $q$.

\textit{Small-scale contributions}. In Fig.~\ref{fig:Coulomb_2},
we present the contributions from small scales
to the ${1/N^{2}}$ relaxation rate.
Starting from Eq.~\eqref{eq:Landau_1N2},
let us place ourselves in the limit where the three actions, $\bJ$,
involved in the three-body resonance condition
are close to one another,
and assume that ${ k_{1} , k_{2} \!\simeq\! k }$.
Then, from eq.~{(B1)} in~\cite{Fouvry2022},
we find that ${ \mU^{(1)}_{k_{1} k_{2}} \!\simeq\! k^{2} U_{k}^{2} }$,
where we assumed that computing gradients of the form ${ \p_{J} U_{k_{1}} (J , \Jp) }$
does not lead to any additional scaling dependence with respect to $k$.
Assuming local resonances,
we may follow Eq.~\eqref{eq:Uk_generic}
and impose the scaling ${ U_{k} \!\simeq\! k^{-1} }$.
We then obtain ${ \mU^{(1)}_{k_{1} k_{2}} \!\simeq\! 1 }$ as a function of $k$.
Similarly, starting from eq.~{(B2)} in~\cite{Fouvry2022},
we find ${ \mU^{(2)}_{k_{1} k_{2}} \!\simeq\! k^{-1} }$.
With these two scalings,
we find from Eq.~\eqref{eq:exp_Lambda2} that ${ |\Lambda_{k_{1}k_{2}} (\bJ) |^{2} \!\simeq\! k^{-4} }$.
Injecting this dependence in Eq.~\eqref{eq:Landau_1N2},
and using the definition from Eq.~\eqref{eq:def_R2kk},
we get ${ \mR_{2} (J , k_{1} , k_{2}) \!\simeq\! k^{-3} }$.
Finally, we may perform the sum over ${ k_{1}, k_{2} }$ in Eq.~\eqref{eq:def_R2kk},
with the constraint ${ \Max [|k_{1}| , |k_{2}|] \!=\! k }$,
to obtain ${ \mR_{2} (J , k) \!\simeq\! k^{-2} }$.
This is the scaling that is indeed recovered in Fig.~\ref{fig:Coulomb_2}.
As expected, in that figure, we also recover
that softening accelerates the decay of the contributions
stemming from small scales.

\textit{Computing the kinetic prediction}. In order to evaluate Eq.~\eqref{eq:Landau_1N2},
we need to compute an integral of the form
\begin{equation}
I = \mP \!\! \int_{\Jminus}^{\Jplus} \!\!\!\! \rd J_{1} \, f (J , J_{1}) ,
\label{eq:shape_int_initial}
\end{equation}
where the integration boundaries
${ J_{\pm} \!=\! J_{0} \!\pm\! \sigma_{0} }$
follow from Eq.~\eqref{eq:DF_LOC}.
This integral can be rewritten as
\begin{equation}
I (J) = \mP \bigg[ \!\! \int_{0}^{J - \Jminus} \!\!\!\!\!\!\!\! \rd \Delta \, f(J , J \!-\! \Delta) + \!\! \int_{0}^{\Jplus - J} \!\!\!\!\!\!\!\! \rd \Delta \, f (J , J \!+\! \Delta) \bigg] .
\label{eq:shape_int}
\end{equation}
It subsequently becomes
\begin{align}
I (J) = \!\! \int_{0}^{\Max [\Dm , \Dp]} \!\!\!\!\!\! \rd \Delta \, \bigg( {} & f (J , J \!-\! \Delta) \, \Theta [\Delta \!\leq\! \Dm ] 
\nonumber
\\
+ {} & f (J , J \!+\! \Delta) \, \Theta [\Delta \!\leq\! \Dp] \bigg) ,
\label{eq:reshape_int}
\end{align}
where we introduced the boundaries ${ \Dm \!=\! J \!-\! \Jminus }$
and ${ \Dp \!=\! \Jplus \!-\! J }$.
With this rewriting, the Cauchy principal value has been safely dealt with.
In practice, we finally compute Eq.~\eqref{eq:reshape_int}
using the midpoint rule with the fixed step distance ${ \Delta_{\rs} \!=\! (\Jplus \!-\! \Jminus) / N_{\rs} }$
with ${ N_{\rs} \!=\! 1\,000}$.
In practice, we truncated the sum over harmonics to ${ |k_{1}| , |k_{2}| \!\leq\! 100 }$
in Eq.~\eqref{eq:Landau_1N2}.

\section{Numerical simulations}
\label{app:NumericalSimulations}

\textit{Equations of motion}. Following Eq.~\eqref{eq:def_Htot_tambouille},
the equations of motion of the point vortices are given by
\begin{equation}
\frac{\rd \br_{i}}{\rd t} = \frac{\rd \br_{i}}{\rd t} \bigg|_{\ext} + \frac{\rd \br_{i}}{\rd t} \bigg|_{\self} .
\label{eq:EOM_split}
\end{equation}
Here, ${ \rd \br / \rd t |_{\ext} }$ is the contribution from the background external
potential, ${ \Uext }$, in Eq.~\eqref{eq:def_Htot_tambouille}.
Since it is axisymmetric, following Eq.~\eqref{eq:def_AA},
we can compute ${ \Omega_{i} \!=\! \Omega (J_{i}) \!=\! \rd \Uext / \rd J_{i} }$,
and we find
\begin{subequations}
\begin{align}
\frac{\rd x_{i}}{\rd t} \bigg|_{\ext} {} & = \;\;\; \Omega_{i} \, y_{i} ,
\label{eq:EOM_bg_x}
\\
\frac{\rd y_{i}}{\rd t} \bigg|_{\ext} {} & = - \Omega_{i} \, x_{i} .
\label{eq:EOM_bg_y}
\end{align}
\label{eq:EOM_bg}\end{subequations}

For the self-interaction part, ${ \rd \br_{i} / \rd t |_{\self} }$,
following Eq.~\eqref{eq:def_U_softened},
we obtain
\begin{subequations}
\begin{align}
\frac{\rd x_{i}}{\rd t} \bigg|_{\self} {} & = - \frac{G}{2 \pi} \sum_{j \neq i} \gamma_{j} \, \frac{y_{i} \!-\! y_{j}}{(x_{i} \!-\! x_{j})^{2} \!+\! (y_{i} \!+\! y_{j})^{2} \!+\! \eps^{2}} ,
\label{eq:EOM_self_x}
\\
\frac{\rd y_{i}}{\rd t} \bigg|_{\self} {} & = \;\;\; \frac{G}{2 \pi} \sum_{j \neq i} \gamma_{j} \, \frac{x_{i} \!-\! x_{j}}{(x_{i} \!-\! x_{j})^{2} \!+\! (y_{i} \!-\! y_{j})^{2} \!+\! \eps^{2}} .
\label{eq:EOM_self_y}
\end{align}
\label{eq:EOM_self}\end{subequations}
In practice, we fixed the softening length to ${ \eps / J_{0} \!=\! 0.01 }$
in all the simulations considered.
The present system has two global invariants,
the total energy, ${ \Etot \!=\! \Htot }$,
as well as the total momentum, ${ \Ltot \!=\! \sum_{i = 1}^{N} \! \gamma_{i} \, r_{i}^{2} }$.
We emphasise that evaluating Eq.~\eqref{eq:EOM_self}
is costly, since it requires to perform ${ \mO (N^{2}) }$ computations
per evaluation of the rates of change.

\textit{Time integration}. The Hamiltonian from Eq.~\eqref{eq:def_Htot_tambouille}
is not under a separable form.
As such, one cannot readily use standard splitting methods
to devise symplectic integration~\cite[see, e.g.\@][]{Hairer+2006}.
This could be circumvented by appropriately ``duplicating phase space''~\citep[see, e.g.\@,][]{Tao2016}.
Though, such symplectic approaches typically require
one to carefully tune some ``binding constant'' between the two copies
of the system's Hamiltonian.
For a given problem, the optimal value of this ad hoc constant,
noted $\omega$ in~\cite{Tao2016},
can depend, among other things, on the value of $N$
or the considered active fraction, $q$.
In the present context, since we are exploring a large range of values in $N$,
and since both the external potential and the pairwise interaction potentials are smooth,
it appears more convenient to resort to more traditional high-order explicit Runge--Kutta methods.
In particular, this allows for the use of a large individual timestep
making long-time integrations easier to perform.
This also removes the need for any additional control parameter to be calibrated.
In that view, we used the 15-stage 9th-order ``more efficient''
\texttt{Vern9} method available through the library \texttt{OrdinaryDiffEq.jl}.

\textit{Simulation setups}. Given that we are going to perform
numerical simulations over some large ranges in $N$ and $q$,
it is important to carefully pick the parameters of the simulations.

\begin{figure}
\begin{center}
\includegraphics[width=0.45\textwidth]{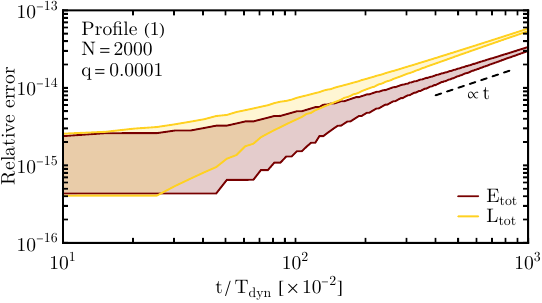}
\includegraphics[width=0.45\textwidth]{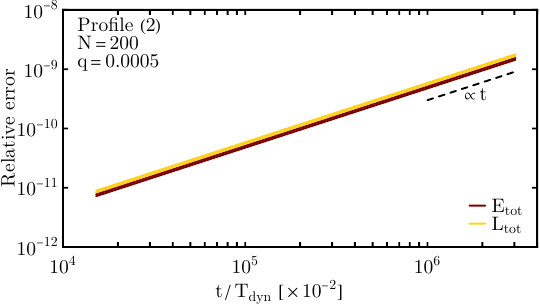}
\caption{Time evolution of the relative error in $\Etot$ and $\Ltot$
for profile (1) (top panel) and profile (2) (bottom panel)
during the $N$-body integration.
The fillings illustrate the spread between the 16\% and 84\% level lines
over the available realisations.
\label{fig:Error}}
\end{center}
\end{figure}

For profile (1), we fixed the timestep to ${ h / \Tdyn \!\simeq\! 1.41 \!\times\! 10^{-2} }$.
Each simulation was integrated up to ${ \Tmax / \Tdyn \!\simeq\! 1.013 \!\times\! 10^{3} \, (0.0001/q)^{2} \, (N/2000) }$,
to match the scaling of the relaxation time expected from Eq.~\eqref{eq:Trelax_1N}.
We made 2\,000 dumps of each realisation as the integration was performed.
The final relative errors on $\Etot$ and $\Ltot$
were typically of order $10^{-13}$,
as illustrated in Fig.~\ref{fig:Error}.
In that same figure, we recover that our explicit integration scheme
leads to a linear growth in time of the error in the global $N$-body invariants.
In practice, the case ${ (N,q) \!=\! (2000, 0.0001) }$ presented in Fig.~\ref{fig:Landau_1}
required ${ \!\sim\! 40 \mathrm{min} }$ of computation time per realisation, on a single core.
Of course, simulating the case ${ (N,q) \!=\! (8\,000, 0.0001) }$,
explored in Fig.~\ref{fig:Treg_Tnl},
was even more numerically costly,
and required ${ \!\sim\! 36\rh }$ of computation time per realisation.

For profile (2), we fixed the timestep to ${ h / \Tdyn \!\simeq\! 1.90 \!\times\! 10^{-2} }$.
Each simulation was integrated up to
${ \Tmax / \Tdyn \!\simeq\! 3.040 \!\times\! 10^{6} \, (0.0005/q)^{4} (N/200)^{2} }$,
to match the scaling from Eq.~\eqref{eq:Trelax_1N2}.
Similarly, we made 2\,000 dumps of each realisation as the integration was performed.
For that profile, the final relative errors on $\Etot$ and $\Ltot$
were of order $10^{-9}$, as illustrated in Fig.~\ref{fig:Error}.
In practice, the case ${ (N,q) \!=\! (200,0.0005) }$ presented
in Fig.~\ref{fig:Landau_2} required ${ \!\sim\! 21\rh }$ of computation time,
per realisation, on a single core.

\begin{figure}
\begin{center}
\includegraphics[width=0.45\textwidth]{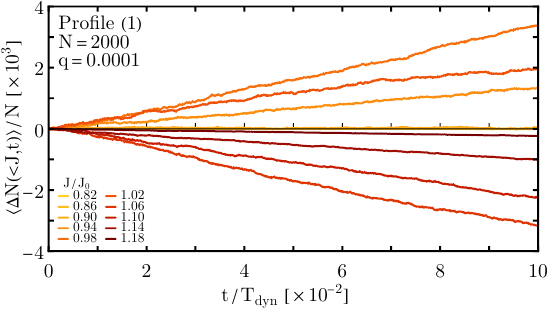}
\includegraphics[width=0.45\textwidth]{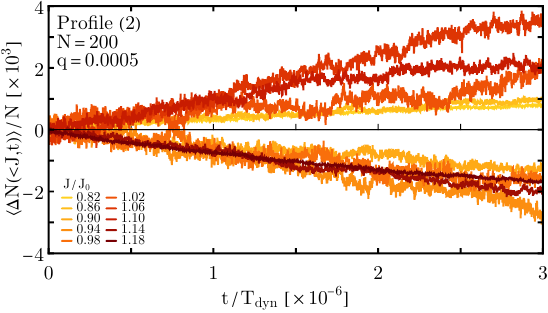}
\caption{Time evolution of the number of particles, ${ N( \!<\!J , t) }$,
in numerical simulations, when averaged over all available realisations,
for profile (1) [top] and profile (2) [bottom].
Different colours correspond to different actions $J$.
The linear dependence of these time series
is the sign of the onset of relaxation.
\label{fig:Nleft_time}}
\end{center}
\end{figure}

\textit{Measuring relaxation}. To obtain the numerical results presented throughout
the main text, we proceeded as follows.
For each realisation, we sliced the action domain,
${ 0.8 \!\leq\! J/J_{0} \!\leq\! 1.2 }$ in 50 locations,
uniformly separated.
Then, for each such action, and each dump, we computed the instantaneous
number of particles with action smaller than the considered action.
Doing so, for each realisation and each considered action, we had at our disposal
a time series of ${ t \!\mapsto\! N ( \!<\! J , t) }$,
as introduced in Eq.~\eqref{eq:def_R1}.
For each value of ${ (N,q) }$ considered,
we ran a total of ${ \Nreal \!=\! 4\,096 }$ independent realisations
in order to estimate ensemble averages.
In Fig.~\ref{fig:Nleft_time}, we illustrate the typical change
in the number of particles, ${ \Delta N( \!<\! J , t) \!=\! N( \!<\! J , t) \!-\! N( \!<\! J , t \!=\! 0) }$,
when averaged over all available realisations.
In that figure, we note that the relative change in the fraction of particles
is of order $10^{-3}$,
i.e.\ we purposely limited our measurements to the onset of relaxation.
Nonetheless, such a relative change is much larger
than the typical relative error in the $N$-body invariants (see Fig.~\ref{fig:Error}).
This strengthens our confidence in the appropriate convergence
of the $N$-body measurements of the relaxation rate.

Once the average time series ${ t \!\mapsto\! \langle N( \!<\! J , t ) \rangle }$ determined,
we performed a linear fit of the form
\begin{equation}
\langle N( \!<\! J , t ) \rangle \simeq \alpha + \beta t .
\label{eq:linear_fit}
\end{equation}
Finally, we use the slope $\beta$ in Eqs.~\eqref{eq:def_R1}
and~\eqref{eq:def_R2} to estimate the relaxation rates.
Given the computing cost associated with each such realisation,
the number of independent realisations,
as well as the large number of values of ${ (N,q) }$ considered,
obtaining Fig.~\ref{fig:Treg_Tnl} was no light numerical endeavour.
Finally, to obtain Fig.~\ref{fig:Relaxation_1}, we binned the action domain
in bins of width ${ 10^{-2} \, J_{0} }$.

\textit{Bootstraps}. In practice, it is essential to estimate the uncertainties associated
with these measurements. To that purpose, we use bootstraps.
A given bootstrap proceeds as follows.
For every value of ${ (N,q) }$,
we perform the ensemble average of ${ N( \!<\! J , t ) }$
over a sample of $\Nreal$ realisations drawn,
with repetitions, from the $\Nreal$ realisations available.
Using the associated ${ \langle N( \!<\! J , t ) \rangle }$,
we perform the linear fit from Eq.~\eqref{eq:linear_fit}
from which we also estimate the variance of the slope $\beta$.
Finally, for each ${ (N,q) }$ and each considered action,
we can draw a value of $\beta$ according to the associated Gaussian distribution.
In practice, we performed a total of 1\,000 bootstrap measurements,
with which we could obtain the 16\% and 84\% level contours
presented in Figs.~\ref{fig:Landau_1} and~\ref{fig:Landau_2},
as well as the median values presented in Figs.~\ref{fig:Flux_Grid_1} and~\ref{fig:Flux_Grid_2}.

\begin{figure}
\begin{center}
\includegraphics[width=0.45\textwidth]{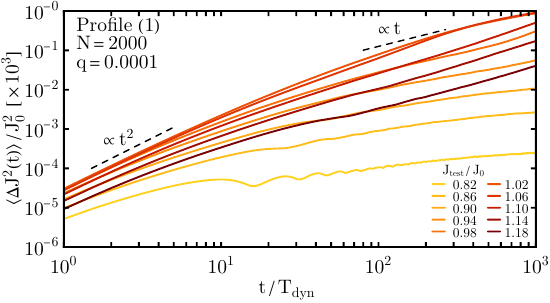}
\caption{Time evolution of the (squared) change in action,
${ \langle \Delta J^{2} (t) \rangle }$, of test particles
in numerical simulations,
when averaged over all available realisations,
for profile (1).
Different colours correspond to different initial actions,
$\Jtest$, for the test particles.
The diffusion of the test particles shows
the transition from a ballistic regime (${ \propto\! t^{2} }$) at early times
to a diffusive one (${ \propto\! t }$) at late times.
\label{fig:DeltaJSQ_time_1}}
\end{center}
\end{figure}

\textit{Diffusion coefficients}.
In order to measure the diffusion coefficients
in Figs.~\ref{fig:Landau_1}, \ref{fig:DeltaJSQ_T},
we proceeded as follows.
We start from the exact same setup
as in the previous $N$-body runs,
using ${ N \!=\! 2\,000 }$ and ${ q \!=\! 0.0001 }$.
We add to each simulation 1\,000 zero-mass test particles,
which, therefore, do not contribute the system's instantaneous potential.
Initially, these particles are all sampled with the same initial action,
${ \Jtest }$, and with their initial angle, $\theta$,
drawn uniformly within ${ [0 , 2 \pi] }$.
In practice, we considered sets of initial conditions
with ${ 0.8 \!\leq\! \Jtest / J_{0} \!\leq\! 1.2 }$
in increments of $ 10^{-3} \, J_{0}$.
For each such choice of $\Jtest$,
we performed 512 independent realisations.
Finally, using the same integration parameters as previously
(timestep, number of dumps, etc.),
for each realisation, we record the time series of ${ t \!\mapsto\! \langle (J (t) \!-\! J (0) )^{2} \rangle }$, averaged over all the test particles from that realisation.
In Fig.~\ref{fig:DeltaJSQ_time_1}, we illustrate
such a time series,
once averaged over all available realisations.
In that figure, we note a transition
between an initial ``ballistic'' regime,
during which ${ \langle \Delta J^{2} (t) \rangle }$
grows like $t^{2}$,
and a ``diffusive'' regime during which it only grows like ${t}$.
In Fig.~\ref{fig:DeltaJSQ_time_1},
we find that the typical relative change in ${ \langle \Delta J^{2} \rangle / J_{0}^{2} }$
at the end of the integration is, at minimum, of order $10^{-7}$.
This is much larger than $10^{-13}$,
the typical relative error in the $N$-body invariants (see Fig.~\ref{fig:Error}).
This brings some confidence in the appropriate convergence
of the $N$-body measurements of the diffusion coefficients.

In order to measure the diffusion coefficient,
we estimate the numerical diffusion coefficient, $\Dnum$,
through the slope
\begin{equation}
\Dnum (\Jtest) = \frac{\langle \Delta J^{2} (\tmax) \rangle - \langle \Delta J^{2} (\tmin) \rangle}{\tmax - \tmax} ,
\label{eq:measure_slope}
\end{equation}
between the times $\tmin$ and $\tmax$.
For Fig.~\ref{fig:Landau_1},
in order not to be polluted by the initial ballistic regime (Fig.~\ref{fig:DeltaJSQ_time_1}),
we considered the time range
${ \tmin \!=\! 500 \, \Tdyn }$ and ${ \tmax \!=\! 1\,000 \, \Tdyn }$.
We checked that varying slightly these values
did not lead to any significant change in the measured $\Dnum$.
For the finite-time diffusion coefficients presented in Fig.~\ref{fig:DeltaJSQ_T},
we fixed ${ \tmin \!=\! 0 }$ and used the values of $\tmax$
indicated in that figure.
In both figures, uncertainties are estimated
using bootstrap resamplings.
This amounts to performing 1\,000 different ensembles averages
over sets of 512 realisations,
but allowing for repetitions.
In Figs.~\ref{fig:Landau_1} and~\ref{fig:DeltaJSQ_T},
we present the 16\% and 84\% level contours
over these bootstraps.

\section{Properties of the regularised Landau equation}
\label{app:reglog}

In this Appendix,
we study the properties of the regularised Landau equation from Eq.~\eqref{eq:Landau_1N_REG}.
In particular, we show how resonance broadening
regularises the divergences of the Landau diffusion coefficient for $2D$ point vortices.

\subsection{Regularisation of the logarithmic divergence of the Landau diffusion coefficient}
\label{app:reg_div}

Let us first consider the case of a monotonic frequency profile,
or a value of $J$ far from the value $\Jstar$,
where the frequency profile has an extremum.

The regularised diffusion coefficient is given by Eq.~\eqref{eq:D2_1N_REG}. Using the expression of the unsoftened potential of interaction from Eq.~\eqref{eq:Uk_generic_full} and the identity ${ \sum_{k>0} x^k / k \!=\! - \ln (1 \!-\! x) }$, we can explicitly perform the sum over $k$ in the function ${ \Lambdatot (J,J_1) }$ \cite{Chavanis2012Vortex}, and obtain
\begin{align}
D(J) = - \tfrac{1}{2} \gamma \!\! \int_0^{+\infty} \!\!\!\!\!\! \rd J_1 \, \ln\left (1-\frac{J_{\mathrm{min}}}{J_{\mathrm{max}}}\right ) \nonumber
\\
\times \,\delta_{\Treg}\left \lbrack \Omega(J)-\Omega(J_1) \right \rbrack F(J_1) ,
\label{eq:r1}
\end{align}
with ${ J_{\mathrm{min}} \!=\! \Min[J , J_{1}] }$ and similarly for $J_{\mathrm{max}}$.

When ${ \Treg \!\rightarrow\! +\infty }$, the Lorentzian tends to a Dirac delta imposing
${ \Omega(J_1) \!=\! \Omega(J) }$, hence ${ J_1 \!=\! J }$.
In that case, the (Landau) diffusion coefficient diverges logarithmically
at small scales, i.e.\ for ${ k \!\rightarrow\! +\infty }$.
Introducing a cut-off $\kmax$,
and using Eq.~\eqref{eq:solve_resonance_condition},
we obtain~\cite{Chavanis2012Vortex}
\begin{align}
D(J) = \tfrac{1}{2}\gamma \frac{F(J)}{|\Omega'(J)|} \ln\Lambda,
\label{eq:r2}
\end{align}
with ${ \ln\Lambda \!=\! \sum_{k>0} \frac{1}{k}\sim \ln k_{\mathrm{max}} }$. 
When $\Treg$ is finite, there is no divergence in Eq.~\eqref{eq:r1}.

Let us now assume that $\Treg$ is sufficiently large.
In that case, the regularised Dirac delta (Lorentzian) is strongly peaked around ${ J_1 \!=\! J }$.
We can therefore make a local approximation and expand ${ \Omega (J_1) }$
to first order around $J$.
We can also replace ${ F (J_1) }$ by ${ F(J) }$.
Doing so, at leading order, we obtain
\begin{equation}
D(J) \sim - \tfrac{1}{2}\gamma F(J) J
\!\! \int_{-\infty}^{+\infty} \!\!\!\!\!\! \rd x \, \ln |x| \, \delta_{\Treg}\left\lbrack \Omega' (J)Jx\right \rbrack,
\label{eq:r3}
\end{equation}
giving
\begin{equation}
D(J) \sim \tfrac{1}{2}\gamma \frac{F(J)}{|\Omega'(J)|} \ln \left \lbrack \Treg |\Omega'(J)|J\right \rbrack.
\label{eq:r4}
\end{equation}
We thus find that the regularized Coulomb logarithm, ${ \ln\Lambda }$,
is equal to the logarithm of the ratio between $\Treg$ and the timescale ${ 1/(|\Omega'(J)|J) }$
set by the shear of the frequency profile.

\subsection{Regularisation of the Landau diffusion coefficient when ${ J \!=\! \Jstar }$}
\label{app:reg_div_Jstar}

We now consider the case where the frequency profile
has a minimum at $\Jstar$, i.e.\ one has ${ \Omega'_{\star} \!=\! \Omega' (\Jstar) \!=\! 0 }$.
In that case, the Landau diffusion coefficient diverges at ${ J \!=\! \Jstar }$,
as discussed in the main text (see also Eq.~\ref{eq:r2}).
Resonance broadening, which is here accounted for heuristically
by a finite value of $\Treg$, regularises this divergence.
Let us consider the function ${ \Dstar(\Treg) }$ obtained by taking ${ J \!=\! \Jstar }$ in Eq.~\eqref{eq:r1}.

For ${ \Treg \!\gg\! \Tregc }$
(with the transition timescale $\Tregc$ to be introduced in Eq.~\ref{eq:r10}),
and using the same method as above but pushing the Taylor expansion up to second order
(assuming ${ \Omega''_{\star} \!=\! \Omega''(\Jstar) \!\neq\! 0 }$),
we find at leading order
\begin{equation}
\Dstar\sim - \tfrac{1}{2}\gamma \Fstar \Jstar
\!\! \int_{-\infty}^{+\infty} \!\!\!\!\!\! \rd x \, \ln|x| \, \delta_{\Treg} \left\lbrack \tfrac{1}{2}\Omega_{\star}'' \Jstar^2x^2\right \rbrack,
\label{eq:r5}
\end{equation}
giving
\begin{equation}
\Dstar\sim \tfrac{1}{8}\gamma \Fstar \sqrt{\frac{\Treg}{| \Omega_{\star}'' |}}\left\lbrace \pi+2\ln \left \lbrack \tfrac{1}{2}|\Omega_{\star}'' |\Jstar^2 \Treg \right \rbrack\right\rbrace.
\label{eq:r6}
\end{equation}
Ignoring logarithmic corrections, and using $\Gamma\sim \Fstar \Jstar$, we obtain the scaling 
\begin{equation}
\Dstar\sim \gamma\Gamma \sqrt{\frac{\Treg}{\Jstar^2 \, | \Omega_{\star}'' |}}\propto \Treg^{1/2}.
\label{eq:r7}
\end{equation}
This scaling remains valid for a softened potential.

For ${ \Treg \!\ll\! \Tregc }$, using ${ \delta_{\Treg} (\omega) \!\rightarrow\! \Treg / \pi }$, we obtain
\begin{align}
\Dstar\sim - \tfrac{1}{2}\gamma \frac{\Treg}{\pi} \!\! \int_0^{+\infty} \!\!\!\!\!\! \rd J_1 \, \ln\left (1-\frac{J_{\mathrm{min}}}{J_{\mathrm{max}}}\right ) F(J_1),
\label{eq:r8}
\end{align}
yielding the scaling 
\begin{align}
\Dstar\sim \gamma\Gamma \Treg \propto \Treg.
\label{eq:r9}
\end{align}
Once again, this scaling remains valid for a softened potential.
The transition between these two regimes occurs at
\begin{align}
\Tregc \sim \frac{1}{\Jstar^2 \, | \Omega_{\star}'' |},\qquad \Dstarc \sim \frac{\gamma \Gamma}{\Jstar^2 \, | \Omega_{\star}'' |}.
\label{eq:r10}
\end{align}
We note that the value of $\Tregc$ is independent of $N$. Since $\Treg$ increases with $N$,
we expect that we are generically in the regime ${ \Treg \!\gg\! \Tregc }$,
not in the regime ${ \Treg \!\ll\! \Tregc }$.

\subsection{Self-consistent relation}

In the main text, $\Treg$ was treated as an adjustable parameter,
see e.g.\@, Fig.~\ref{fig:Flux_Treg_1}.
Now, if we assume the relation
\begin{equation}
\Treg \sim\frac{\Jstar^2}{\Dstar},
\label{eq:r11}
\end{equation}
and use Eq.~\eqref{eq:r1} with ${ J \!=\! \Jstar }$,
we can obtain the expression of $\Treg$ (or $\Dstar$) self-consistently.
We refer to Appendix~\ref{app:t} for a graphical study.

For ${ \Treg \!\gg\! \Tregc }$ (i.e.\ for ${ \Dstar \!\gg\! \Dstarc }$),
combining Eqs.~\eqref{eq:r7} and~\eqref{eq:r11}, we get
\begin{equation}
\Dstar \sim \left (\frac{\gamma\Gamma}{\sqrt{| \Omega_{\star}'' |}}\right )^{2/3},\quad \Treg \sim \Jstar^2\left (\frac{\sqrt{| \Omega_{\star}'' |}}{\gamma\Gamma}\right )^{2/3} .
\label{eq:r12}
\end{equation}
This gives ${ \Treg \!\propto\! \Dstar^{-1} \!\propto\! \gamma^{-2/3} \!\propto\! N^{2/3} }$. To the best of our knowledge, such a $N^{2/3}$ scaling has not been reported before.

For ${ \Treg \!\ll\! \Tregc }$ (i.e.\ for ${ \Dstar \!\ll\! \Dstarc }$),
combining Eqs.~\eqref{eq:r9} and~\eqref{eq:r11}, we get
\begin{align}
\Dstar \sim \Jstar\sqrt{\gamma\Gamma},\quad \Treg \sim \frac{\Jstar}{\sqrt{\gamma\Gamma}},
\label{eq:r13}
\end{align}
giving ${ \Treg \!\sim\! \Dstar^{-1} \!\sim\! \gamma^{-1/2} \!\sim\! N^{1/2} }$.
This is the $N^{1/2}$ scaling presented in the main text,
which implicitly assumed ${ \Treg \!\ll\! \Tregc }$ (see Section~\ref{sec:1N_ResonanceBroadening}).
It corresponds to the Taylor-McNamarra~\cite{Taylor+1971} scaling
which applies to an unsheared flow (see Appendix~\ref{app:link_previous}).
In the case considered in~\cite{Taylor+1971},
${ \Omega(J) }$ is constant, so that ${ \Omega''(J) }$ vanishes everywhere,
as well as all its higher order derivatives.
This implies ${ \Tregc \!\rightarrow\! +\infty }$ and ${ \Dstarc \!\rightarrow\! 0 }$,
following Eq.~\eqref{eq:r10}.
By contrast, when ${ \Omega_{\star}'' \!\neq\! 0 }$, as in our case,
there exists another regime, ${ \Treg \!\gg\! \Tregc }$,
that yields a $N^{2/3}$ scaling.
It could be that this scaling,
which is valid when ${ |\Omega_{\star}''| \!\gg\! \sqrt{\gamma\Gamma} / \Jstar^{3} \!\sim\! 1/\sqrt{N} }$,
is more relevant than the $N^{1/2}$ scaling,
valid in the opposite limit,
presented in the main text.
Unfortunately, these two scalings cannot be distinguished in the numerical simulations
presented here, since $2/3$ is close to $1/2$.
This subtle distinction will be considered in a future work,
by studying a simpler system
for which more precise numerical simulations can be performed. 

\textit{Remark}. One can generalise the previous results to the case
where the first ${ n \!-\! 1 }$ derivatives of ${ \Omega(J) }$
vanish at $\Jstar$
but ${ \Omega_{\star}^{(n)} \!=\! \Omega^{(n)} (\Jstar) \!\neq\! 0 }$.
Instead of Eq.~\eqref{eq:r7}, we find
\begin{equation}
\Dstar\sim \frac{\gamma\Gamma \Treg^{1-1/n}}{\Jstar \big| \Omega_{\star}^{(n)} \big|^{1/n}}
\label{eq:r14}
\end{equation}
for ${ \Treg \!\gg\! 1 / (\Jstar^{n} | \Omega_{\star}^{(n)} |) }$.
Then, instead of Eq.~\eqref{eq:r12}, we find
\begin{equation}
\Treg \sim\frac{\Jstar^2}{\Dstar}\sim \left\lbrack \frac{\Jstar^3 \big| \Omega_{\star}^{(n)} \big|^{1/n}}{\gamma\Gamma}\right \rbrack^{\frac{1}{2-1/n}}
\label{eq:r15}
\end{equation}
for ${ |\Omega_{\star}^{(n)} | \!\gg\! \sqrt{\gamma\Gamma}/\Jstar^{n+1} \!\sim\! 1/\sqrt{N} }$.
This yields the general scaling ${ \Treg \!\sim\! \Dstar^{-1} \!\sim\! \gamma^{-1/(2-1/n)} \!\sim\! N^{1/(2-1/n)} }$.
The results from Eqs.~\eqref{eq:r8} and~\eqref{eq:r13},
corresponding to the Taylor--McNamara regime~\citep{Taylor+1971}, are not changed.
For ${ n \!=\! 1 }$ we recover the Landau scaling,
for ${ n \!=\! 2 }$ we recover the new scaling from Eq.~\eqref{eq:r12},
and for ${ n \!\rightarrow\! +\infty }$ we recover the Taylor--McNamara~\citep{Taylor+1971} scaling.
Exploring these different regimes of relaxation
will be the topic of a future work.

\section{Resonance broadening from stochastic equations of motion}
\label{app:t}

In this Appendix, we introduce an alternative stochastic model
to justify the heuristic resonance broadening from Section~\ref{sec:1N_ResonanceBroadening}
and motivate the regularisation of the Dirac delta by a Lorentzian function (Eq.~\ref{eq:def_Lorentzian}).
In addition, we use this model to estimate the diffusion coefficient of point vortices.
We also point out the limitations of this approach
and the need for a more refined kinetic framework.
Finally, we discuss the connection between this work
and previous literature.

\subsection{Regularised Landau equation from individual stochastic equations}
\label{app:Landau_from_stochastic}

When derived from the BBGKY hierarchy at order ${ 1/N }$,
the Landau equation generically reads~\cite{Chavanis2013}
\begin{align}
\frac{\p F (J , t)}{\p t} = {} & \frac{\p}{\p
J} \bigg[ \!\int_0^{+\infty} \!\!\!\!\!\! \rd \tau \, \! \int_{0}^{+\infty} \!\!\!\!\!\! \rd J_1 \, \!\! \int_0^{2\pi} \!\!\!\! \rd \theta_1 \, 
\nonumber
\\
{} & \times \big\langle \mF (1\rightarrow
0,t) \, G(t,t-\tau) \, \mF (1\rightarrow
0,t-\tau) \big\rangle
\nonumber
\\
{} & \times \bigg( \frac{\p}{\p
J}-\frac{\p}{\p J_1} \bigg) F(J,t) \, \frac{1}{ \gamma} F(J_1,t) \bigg],
\label{eq:t1}
\end{align}
where 
\begin{equation}
\mF (1\rightarrow
0,s) = - \ri \gamma\sum_{k} k \, \re^{\ri k [
\theta(s) - \theta_1 (s)]} \, U_{k}(J,J_1) 
\label{eq:t2}
\end{equation}
is the force (velocity) created by particle (vortex) $1$ on particle (vortex)
$0$ at time $s$, and ${ G(t , t \!-\! \tau) }$ is an appropriate Green function taking
into account the motion of the particles between $t$ and ${ t \!-\! \tau }$~\cite{Chavanis2013}.
In Eq.~\eqref{eq:t1}, the brackets
indicate an average over the fluctuating trajectories of the particles (see
below). To calculate the temporal correlation function in Eq.~\eqref{eq:t1},
one usually assumes
that the particles follow the unperturbed mean field trajectories at leading order.
This leads to the Landau equation (Eq.~\ref{eq:Landau_1N_noK}).
However, this approximation breaks down notably
close to an extremum of ${ \Omega (J) }$,
i.e.\ when particles have similar frequencies.
In that case, one has to take into account fluctuations
in the particles' individual trajectories.

Following~\cite{Dubin2003}, we replace the deterministic equations of motion
by effective stochastic equations where the amplitude
of the noise term is determined by the diffusion coefficient
obtained from the kinetic theory itself (see below).
Assuming that the diffusion is isotropic~\cite{Dubin2003},
we then have ${ D_{\theta\theta}(J) \!=\! D(J) / (4J^2) }$,
where ${ D(J) \!\equiv\! D_{JJ}(J) }$ [resp.\ ${ D_{\theta\theta} (J) }$]
is the diffusion coefficient in action [resp.\ angle].

Since our goal is to illustrate the origin of resonance broadening
in a simple manner,
we make the following approximations:
(i) We assume that the stochastic process can be described by a Gaussian white noise;
(ii) We ignore the diffusion in action in the stochastic equations of motion;
(iii) We assume that the diffusion coefficient in angle, $D_{\theta\theta}$,
is constant with value ${ \Dstar / (4 \Jstar^{2}) }$,
where ${ \Dstar \!=\! D (\Jstar) }$ is the action diffusion coefficient
at the extremum of the frequency profile.
As a result, we approximate the motion of the particles
with the effective stochastic equations 
\begin{subequations}
\begin{align}
\frac{\rd J}{\rd t} {} & = 0,
\label{eq:t3_J}
\\
\frac{\rd \theta}{\rd t} {} & = \Omega(J) + \sqrt{\Dstar / (4 \Jstar^{2})} \, \eta(t) , 
\label{eq:t3_theta}
\end{align}
\label{eq:t3}\end{subequations}
where ${ \eta (t) }$ is a Gaussian white
noise such that ${ \langle \eta(t) \rangle \!=\! 0 }$ and
${ \langle \eta(t) \, \eta(t') \rangle \!=\! \deltaD (t \!-\! t') }$.
Integrating these equations between $t$ and ${ t \!-\! \tau }$,
we obtain the stochastic trajectories
\begin{subequations}
\begin{align}
J (t \!-\! \tau) {} & = J ,
\label{eq:stoch_J_sol}
\\
\theta (t \!-\! \tau) {} & = \theta \!-\! \Omega (J) \tau \!-\! \sqrt{\Dstar / (4 \Jstar^{2})} \! \int_{0}^{\tau} \!\! \rd t' \, \eta (t \!-\! t') .
\end{align}
\label{eq:stoch_sol}\end{subequations}
Substituting these solutions into Eq.~\eqref{eq:t1}
and carrying out the integration over $\theta_{1}$,
we obtain
\begin{align}
{} & \frac{\p F (J,t)}{\p t} = 4\pi\gamma \frac{\p}{\p
J} \bigg[ \!\! \int_{0}^{+\infty} \!\!\!\!\!\!\!\!\! \rd J_1 \, \!\! \int_0^{+\infty} \!\!\!\!\!\!\!\!\! \rd \tau
\sum_{k>0} k^2 \, 
|U_{k}(J,J_1)|^2
\nonumber
\\
{} & \quad \times \re^{\ri k \Delta \Omega (J, J_{1}) \tau} \left\langle
\re^{\ri k \sqrt{\Dstar / (4 \Jstar^{2})} \! \int_0^\tau \! \rd t' [
\eta(t-t')-\eta_1(t-t') ]}
\right\rangle
\nonumber
\\
{} & \quad \times \bigg( \frac{\p}{\p
J}-\frac{\p}{\p J_1} \bigg) F(J,t) F(J_1,t) \bigg] ,
\label{eq:t4}
\end{align}
where we introduced the frequency difference, ${ \Delta \Omega (J , J_{1}) \!=\! \Omega (J) \!-\! \Omega (J_{1}) }$.
Now, if we assume a Gaussian distribution for the fluctuations,
we have~\cite{Dubin2003}
\begin{equation}
\left\langle
\re^{ \ri k \sqrt{\Dstar / (4 \Jstar^{2})} \!\int_0^\tau \! \rd t' \eta(t-t')} \right\rangle = \re^{- k^2 \Dstar / (8 \Jstar^{2})\tau},
\label{eq:t5}
\end{equation}
and the kinetic equation (Eq.~\ref{eq:t1}) becomes
\begin{align}
{} & \frac{\p F (J,t)}{\p t} = 4\pi \gamma \frac{\p}{\p
J} \bigg[ \! \int_{0}^{+\infty} \!\!\!\!\!\!\!\!\! \rd J_1 \!\! \int_0^{+\infty} \!\!\!\!\!\!\!\!\! \rd \tau \sum_{k>0} k^2 |U_{k}(J,J_1)|^2
\label{eq:t6}
\\
{} & \quad \times \re^{\ri k \Delta \Omega (J , J_{1}) \tau} \re^{-\tau/T_{k}} \bigg( \frac{\p}{\p
J}-\frac{\p}{\p J_1} \bigg) F(J,t)F(J_1,t) \bigg] ,
\nonumber
\end{align}
where we introduced the (scale-dependent) decorrelation time
\begin{equation}
T_{k} \!=\! \frac{4J_{\star}^2}{D_{\star} k^2}.
\label{eq:tk}
\end{equation}

Integrating Eq.~\eqref{eq:t6} over $\tau$, we obtain
\begin{align}
{} & \frac{\p F (J,t)}{\p t} = 4\pi^2 \gamma \frac{\p}{\p
J} \bigg[ \!\int_{0}^{+\infty} \!\!\!\! \rd J_1 \sum_{k>0} k^2 \,
|U_{k}(J,J_1)|^2
\label{eq:t7}
\\
{} & \times \delta_{T_{k}} \big( k \, \Delta \Omega [J , J_{1}] \big)
\bigg(\frac{\p}{\p
J}-\frac{\p}{\p J_1} \bigg) F(J,t)F(J_1,t) \bigg] ,
\nonumber
\end{align}
with ${ \delta_{T_{k}} (\omega) }$ the broadened Dirac delta
from Eq.~\eqref{eq:def_Lorentzian}.\footnote{In the main text, we implicitly assumed
${ T_{k} \!=\! \Treg / k }$. This leads to the identity
${ \delta_{T_k} (k \, \omega) \!=\! \delta_{\Treg} (\omega) / |k| }$,
similar to the one satisfied by the usual Dirac delta. From this identity results the
broadened Landau equation (Eq.~\ref{eq:Landau_1N_REG}). Here, the
function ${ \delta_{T_k} (k \, \omega) }$ has a different dependence with $k$, so
the broadened Landau equation (Eq.~\ref{eq:t7}) is slightly different.
See~\cite{Chavanis+2007} for
another regularisation of the Dirac delta obtained by integrating over a finite period of time.}
Explicitly, the kinetic equation reads
\begin{align}
{}& \frac{\p F (J,t)}{\p t}= 4\pi \gamma \frac{\p}{\p
J} \bigg[ \! \int_{0}^{+\infty} \!\!\!\! \rd J_1 \sum_{k>0} k^2 \,
|U_{k}(J,J_1)|^2
\label{eq:t8}
\\
{} & \times \frac{T_{k}}{1\!+\! 
( k \, \Delta \Omega [J , J_{1}] \, T_k )^2}
\bigg( \frac{\p}{\p J} - \frac{\p}{\p J_1} \bigg) F(J,t)F(J_1,t) \bigg].
\nonumber
\end{align}
If we let
${ T_{k} \!\rightarrow\! +\infty }$,
or if ${ k \, \Delta \Omega \, T_k \!\gg\! 1 }$,
which both amount to neglecting the fluctuations in particles individual trajectories,
Eq.~\eqref{eq:t8} becomes the Landau equation (Eq.~\ref{eq:Landau_1N_noK}). This is valid if we are far from the extremum of
${ \Omega(J) }$.

However, when $J$ is close to $\Jstar$,
we are faced with the opposite limit,
namely ${ k \, \Delta\Omega \, T_k \!\ll\! 1 }$.
In that case, the kinetic equation (Eq.~\ref{eq:t8}) reduces to
\begin{align}
\frac{\p F (J,t)}{\p t} = {} & 4\pi \gamma \frac{\p}{\p
J} \bigg[ \! \int_{0}^{+\infty} \!\!\!\! \rd J_1 \sum_{k>0} k^2 \,
|U_{k}(J,J_1)|^2
\nonumber
\\
{} &\times T_{k} \,
\bigg(\frac{\p}{\p
J}-\frac{\p}{\p J_1} \bigg) F(J,t)F(J_1,t) \bigg] .
\label{eq:t9}
\end{align}

\subsection{Diffusion coefficient at the frequency extremum}
\label{app:Diff_extremum}

The (broadened) Landau equation (Eq.~\ref{eq:t7})
can be viewed as a Fokker--Planck
equation in action space involving a diffusion term and a drift term. The diffusion coefficient in action reads
\begin{align}
D(J) = 8 \pi^2 \gamma \!\! \int_{0}^{+\infty} \!\!\!\!\!\! \rd J_1 \sum_{k>0} k^2 \,
|U_{k}(J,J_1)|^2
\nonumber
\\
\times \, \delta_{T_{k}} \big( k \, \Delta \Omega [J , J_{1}] \big) F(J_1,t) .
\label{eq:t7b}
\end{align}
As expected, when ${ T_k \!\rightarrow\! + \infty }$, this equation reduces to the usual Landau diffusion coefficient (Eq.~\ref{eq:def_D2}),
which scales like ${1/N}$ via the prefactor $\gamma$.
This limit is relevant if we are sufficiently far from the extremum of frequency. In that case~\cite{Chavanis+2007,Chavanis2023},
we get
\begin{equation}
D(J) = 8 \pi^2 \gamma \! \sum_{\rr} \! \sum_{k>0} k \,
|U_{k}(J,\Jr)|^2\frac{F(\Jr,t)}{|\Omega'(\Jr)|},
\label{eq:t7bb}
\end{equation}
where $\Jr$ are the actions of the resonant particles such that ${ \Omega(\Jr) \!=\! \Omega(J) }$ (see Eq.~\ref{eq:solve_resonance_condition}).

When ${ J \!\to\! \Jstar }$, following the same argument as in Eq.~\eqref{eq:DL_flux_continued_mD},
the resonance condition has two roots.
Assuming that the frequency profile is quadratic about $\Jstar$,
these two roots are
${ \Jr \!=\! J }$ and ${ \Jr \!=\! 2 \Jstar \!-\! J }$,
which are symmetric with respect to $\Jstar$.
As such, when ${ J \!\to\! \Jstar }$,
we obtain
\begin{equation}
D(J) \simeq 16 \pi^2 \gamma \sum_{k>0} k \,
|U_{k}(\Jstar , \Jstar)|^2\frac{F(\Jstar,t)}{|\Omega'(J)|}.
\label{eq:t7bc}
\end{equation}
We see that the diffusion coefficient diverges at the extremum of frequency,
because of the factor ${ 1/|\Omega'(J)| }$.
This is the same result as the one obtained in Eq.~\eqref{eq:DL_flux_final_mD}.
When accounting for resonance broadening,
this divergence is regularised by the finite value of $T_{k}$,
just like it was regularised by $\Treg$ in Eq.~\eqref{eq:D2_1N_REG},
as visible in Fig.~\ref{fig:Flux_Treg_1}.\footnote{Similarly,
in the absence of softening,
${ \sum_{k} k |U_{k} (\Jstar , \Jstar)|^{2} }$
as present in Eq.~\eqref{eq:t7bc},
diverges logarithmically on small scales (see Fig.~\ref{fig:Coulomb_1}).
Fortunately, given that $T_{k}$ also depends on $k$ (see Eq.~\ref{eq:tk}),
one can show that resonance broadening also regularises
this small scale divergence, see Appendix~\ref{app:reg_div}.}
Unfortunately, the regularisation from Eq.~\eqref{eq:Landau_1N_REG}
in the main text
still involves the adjustable parameter, $\Treg$,
that remains to be determined.
In Appendix~\ref{app:reglog}, we have estimated $\Treg$
by heuristically relating $\Treg$ to $\Dstar$ through Eq.~\eqref{eq:r11}.
However, this equation was not given a precise justification.

In the present approach, $T_{k}$ is related to $\Dstar$
through the similar Eq.~\eqref{eq:tk},
which now arises from a simple stochastic model.
Like in Appendix~\ref{app:reglog}, one could then hope to obtain the value of $T_{k}$
self-consistently through $\Dstar$ (see Eq.~\ref{eq:tk}),
by applying Eq.~\eqref{eq:t7b} at ${ J \!=\! \Jstar }$.
Indeed, this leads to a self-consistent relation of the form
\begin{equation}
\Dstar = f [\Dstar] ,
\label{eq:def_selfC}
\end{equation}
with the function
\begin{align}
f [\Dstar] = 8 \pi^{2} \gamma {} & \!\! \int_{0}^{+ \infty} \!\!\!\!\!\! \rd J_{1} \, \sum_{k > 0} k^{2} | U_{k} (\Jstar , J_{1}) |^{2}
\nonumber
\\
{} & \times \delta_{T_{k}} \big( k \, \Delta \Omega [\Jstar , J_{1}] \big) F(J_1,t) ,
\label{eq:def_f_selfC}
\end{align}
where ${ T_{k} \!=\! T_{k} [\Dstar] }$ follows from Eq.~\eqref{eq:tk}.
In the present heuristic approach,
the self-consistent solution of Eq.~\eqref{eq:def_selfC}
should determine $\Dstar$, hence quantifying the level
of resonance broadening occurring in the system.

\begin{figure}
\begin{center}
\includegraphics[width=0.45\textwidth]{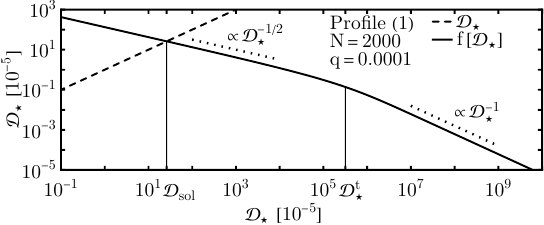}
\caption{Illustration of the self-consistent relation,
${ D_{\star} \!=\! f (D_{\star}) }$ from Eq.~\eqref{eq:def_f_selfC},
for profile (1),
here using the rescaled diffusion coefficient,
$\mD_{\star}$, following Eq.~\eqref{eq:def_mD}.
Here, $\mD_{\star}^{\mathrm{t}}$ illustrates the transition between
the two regimes investigated in Appendix~\ref{app:reg_div_Jstar}.
For the case ${ N \!=\! 2\,000 }$ and ${q \!=\! 0.0001}$,
the fixed-point solution is found for ${ \mDsol \!\simeq\! 2.6 \!\times\! 10^{6} }$.
This is ${ \!\sim\! }$ 20x too large compared
to the $N$-body value estimated in Fig.~\ref{fig:Landau_1}.
\label{fig:Dstar}}
\end{center}
\end{figure}

In Fig.~\ref{fig:Dstar},
we illustrate the function ${ D_{\star} \!\mapsto\! f [D_{\star}] }$,
with the associated fixed-point search.\footnote{This function has the same asymptotic behaviours,
${ f(\Dstar) \!\propto\! \Dstar^{-1/2} }$ for ${ \Dstar \!\ll\! \Dstart }$
and ${ f(\Dstar) \!\propto\! \Dstar^{-1} }$ for ${ \Dstar \!\gg\! \Dstart }$ with ${ \Dstart \!\sim\! \Jstar^2 / \Tregc \!\sim\! \Jstar^4 |\Omega_{\star}'' | }$,
as the equivalent function studied
in Appendix~\ref{app:reglog} with a simpler model.
Therefore, the discussion of Appendix~\ref{app:reglog}
on the different regimes also applies to the present situation.}
In that figure, considering the same case as in Fig.~\ref{fig:Landau_1},
we find the fixed-point solution, ${ \mDsol \!\simeq\! 2.6 \!\times\! 10^{6} }$.
This is ${ \!\sim\! }$ 20x larger than the value for $\mD_{\star}$
effectively measured in the numerical simulations.
This shows the limitation of the simple stochastic model presented in this Appendix.
Surely a refined description of the present resonant broadening is required,
e.g.\@, by considering more intricate
individual stochastic equations of motion (Eq.~\ref{eq:t3}).

\subsection{Link with previous works}
\label{app:link_previous}

The simplified model from Eq.~\eqref{eq:t7} allows us
to draw some connections
between the Landau diffusion coefficient of sheared flows~\cite{Chavanis1998,Chavanis2001,Dubin2003}
and the Taylor--McNamara diffusion coefficient of unsheared flows~\cite{Taylor+1971}.
These correspond to two important limits considered in the literature,
each with different scalings with respect to $N$.

For the sake of simplicity,
let us assume that the vorticity and frequency profiles are approximately flat on a domain of size $\Jstar$, writing ${ F \!\simeq\! \Fstar }$
and assuming ${ k \, \Delta\Omega \, T_k \!\ll\! 1 }$.
From Eq.~\eqref{eq:t7b}, we obtain
\begin{equation}
D^2 \simeq 32\pi\gamma \, \Jstar^2 \Fstar \!\! \int_{0}^{\Jstar} \!\!\!\! \rd J_1 \sum_{k>0}
|U_{k}(J,J_1)|^2.
\label{eq:t10}
\end{equation}
Since ${ \gamma \!\sim\! 1/N }$,
we recover the Taylor--McNamara scaling~\cite{Taylor+1971},
${ D \!\sim\! 1/\sqrt{N} }$.
If we account for an active fraction of particles,
as in the main text,
this scaling
becomes ${ D \!\sim\! q/\sqrt{N} }$.

The integral in Eq.~\eqref{eq:t10} can be estimated analytically
when using the (unsoftened) interaction potential from Eq.~\eqref{eq:Uk_generic_full}.
It leads to
\begin{equation}
D^2 \simeq \frac{2 \Jstar^{3}}{\pi} \, \gamma \, \Fstar ,
\label{eq:t11}
\end{equation}
up to logarithmic corrections.
To make the connection with the results of 
Taylor--McNamara~\cite{Taylor+1971},
we introduce the original isotropic diffusion coefficient in the physical space,
${ (x,y) }$, via ${ D_{\iso} \!\sim\! D / (2J_{\star}) }$, as well as
the total circulation ${ \Gamma \!\sim\! 2\pi \Fstar \Jstar }$.
We obtain
\begin{equation}
D_{\iso}^2 \simeq \frac{\Jstar}{2\pi} \, \gamma \, \Fstar \sim \frac{1}{4\pi^2} \, \Gamma
\gamma \, \sim \frac{1}{4\pi^2} \, N \, \gamma^2 .
\label{eq:t12}
\end{equation}
Of course, following our numerous approximations,
the prefactor of Eq.~\eqref{eq:t12} is uncertain by a factor of order unity.
Nonetheless, this should be compared with the Taylor--McNamara formula, ${ D_{\rTN}^{2} \!=\! N \gamma^2/(16\pi^3) }$~\cite{Taylor+1971}.
Our estimate of $D_{\iso}$ turns out to be larger than
$D_{\rTN}$ by a factor ${ \sqrt{4\pi} \!\sim\! 3.5 }$.

\end{document}